\RequirePackage{fix-cm}
\documentclass[natbib,smallextended]{svjour3}       
\usepackage{graphicx}
\usepackage{caption}
\usepackage{comment}
\usepackage{amsmath}
\usepackage{cases}
\usepackage{geometry}
 \journalname{}
\begin{document}

\title{h-index and its alternative: A Review
}


\author{Anand Bihari         \and
	Sudhakar Tripathi \and Akshay Deepak 
}

\institute{Anand Bihari \at
	Department of Computer Science \& Engineering\\
	National Institute of Technology Patna, Bihar, India
	\email{anand.cse15@nitp.ac.in, csanandk@gmail.com}           
	\and
	Sudhakar Tripathi \at
	Department of Information \& Technology\\
	R. E. C. Ambedkar Nagar, Uttar Pradesh India
	\email{p.stripathi@gmail.com} 
	\and
	Akshay Deepak \at
	Department of Computer Science \& Engineering\\
	National Institute of Technology Patna, Bihar, India
	\email{akshayd@nitp.ac.in}
}
\date{}

\maketitle

\begin{abstract}
In recent years, several Scientometrics and Bibliometrics indicators were proposed to evaluate the scientific impact of individuals, institutions, colleges, universities and research teams. The h-index gives a major breakthrough in the research community to evaluate the  scientific impact of individual. It got a lot of attention due to its simplicity and several other indicators were proposed to extend the properties of h-index as well as to overcome shortcomings of h-index. In this literature review, we have discussed advantages and limitations of almost all Scientometrics as well as Bibliometrics indicators which have been  categorised in eight categories :(i) Complement of h-index, (ii) Based on total number of  authors, (iii) Based on publication age, (iv) Combination of two indices, (v) Based on excess citation count, (vi) Based on total publication count, (vii) Based on other variants and (viii) Based on account of self citation count. The main objective of this article is to study all those indicators which has been proposed to evaluate the scientific impact of individual researcher or a group of researchers.  
\end{abstract}

\section{Introduction}
\label{intro}
Research is a cyclic process to spread  knowledge and innovation in the society for its betterment. The one of the most important product of research is the research articles, that express the knowledge and working process of the proposed model. A scholar first studies the previously deployed methodology. He then looks for ways in which the acceptability of existing methods can be improved or, proposes a new methodology that overcome the limitation(s) of existing ones.  His final outcome is published in the form of a research article. While writing the research article, the scholar refers published research articles. The acceptability of the new technique or research article is measured in terms of citation count (that is, how many research articles refer it) and the impact of the scholar is measured in terms of total citation count earned by all published articles, in which the scholar is present as a author or as a co-author. 

During the last few decades, it seems that the evaluation of the scientific impact of a scholar or a group of scholars was a significant assignment. In many cases, it is compulsory to know the scientific impact of a scholar, for example, at the time of hiring a new faculty member, promotion of faculty members, continuation of research grants, etc. To do this, mostly they used either the total number of publications, total citation count, average citation count or citation count per publication. It is not uncommon to find a scholar who has published high number of articles, but has less scientific impact (i.e, citation count)  than a scholar who has published less number of articles but has a greater scientific impact (citation count). Similarly, a scholar can gain more number of citations with the help of a few articles, however another scholar can gain similar number of citations in a distributed manner. In such a case both scholars are treated equally.  As similar to the total publication and citation count, the average number of citation counts may be reflected by the single highly influenced article. However, the total publication count, the total citation count and the average citation count may not be a good measure, because they do not consider the overall impact and the productivity of scholars, and hence, also fail in comparing the scholars' scientific impact. Thus, we can not use the total publication count, total citation count and average citation count as a measure to assess the scientific impact of scholars. We need a more fine grained approach that consider more number of attributes, because nowadays the research assignment, project grant, faculty promotion, award distribution etc. related decisions are made based on the individual scientific excellence or the performance of the individual in the group of scholars. Several publication indexing databases such as Google Scholar (\cite{walters2007google}), ISI Web of Science, Scopus, DBLP, PubMed, SciELO, CiteSeer, CiteSeerX and many more are available to manage the research publications of scholars. To validate the properties of proposed indicators, scholars use these publication databases (\cite{bar2008informetrics}).

The h-index was proposed by \cite{hirsch2005index}, gives a major breakthrough in the scientific community to assess the scientific impact of scholars and gained a lot of attention due to its simplicity. The h-index covers the productivity and the impact of  scholars and is better than the total number of publications and mean number of citation counts \cite{lehmann2006measures,hirsch2007does}. But the h-index does not consider the impact of excess citation count and leaves a  huge amount of citation count unaccounted. It also fails in comparing the scientific impact of scholars have similar index value. Based on the limitations of the h-index, several other indicators were proposed to overcome the shortcoming of h-index and enhance the scientific evaluation process with more variability. The h-index shows the popularity of a scholar, but it does not mean that the scholar is more prestigious. Because the h-index does not exclude the self citation count and does not differentiate the source of the citation.  The citation counts from different type of articles such as Patents, Journal articles (reputed or non-reputed), conference proceedings and book chapters are treated equally. However, the citation sources and the citation count have an impact on the quality of citation, which is generally not considered \cite{ding2011popular}. 

Nowadays, the increase in collaborations between scholars affect the scientific productivity and the impact of research publications, because, generally,  the research projects are too large and interdisciplinary.  In this case, if we consider the h-index as an assessment tool, then every scholar in multi-authored articles gets full credit of its citation count. However the contribution of all scholars are not equal, so we can not distribute citation equally to all scholars. In this context, several indices were designed to give credit to all scholars and evaluate the scientific impact of scholars based on their credit. \cite{cole1973social} considers only the first author, \cite{multiple1981price} gives equal fractional credit to all scholars, i.e. 1/k, where k is the total number of scholars. \cite{hodge1981publication}, \cite{sekercioglu2008quantifying} and \cite{hagen2010harmonic} gives credit to all scholars as per their proportional rank. \cite{van1997fractional} and \cite{trueba2004robust} used the arithmetic counting for credit allocation, \cite{egghe2000methods} used the geometric series for credit allocation and several other methods are used to distribute shared-credit between all scholars. Most of the indices used the mathematical equation to share the credit among scholars, but we cannot express the contribution of scholars in the form such a mathematical expression. 

In scientific assessment of scholars, the h-index considers only the few highly cited articles. However the articles having at least one citation count have significance in scientific assessment of scholars. In this context \cite{garcia2009multidimensional} designed a multidimensional h-index that covers all cited articles. Several other indices use different mechanism to consider the impact of all cited articles in scientific assessment of scholars.   

It seems that any one indicator does not fulfil all the requirements of scientific assessment process of a scholar. Hence, a combination of two or more different types of indicators is required. The combination of  the properties of two or more different measures is a good step to evaluate the scientific impact of individuals \cite{martin1996use,van2003holy}. In this context hg-index \cite{alonso2010hg}, $q^{2}$-index \cite{cabrerizo2010q2} and the  $h_{mc}$-index \cite{liu2012modifying}) were proposed.

One of the main issues related to most of the proposed indices is that the proposed indices do not consider the publication consistency issue. In some cases, it seems that the some of the scholars have publications only during their doctoral research, post-doctoral research and when working on the research projects. However, some of the scholars publish articles continuously. This issue do not affect the h-index, but has significance in scientific assessment. In this context, the career year h-indices were designed by \cite{mahbuba2012diffusion}.. 
The main objective of this article is to make an extensive literature review on h-index and its variants to focus on following points:
\begin{itemize}
	\item The definition of h-index along with its advantages and limitation.
	\item Literature review on the variants of h-index which are based on different parameters such as total number of citation count, total publication age, total number of collaborators, normalize citation count based on number of co-authors, total number of citers (citing authors) and many more.
\end{itemize}

In this article, section \ref{h-index} presents the definition of the h-index along with its advantages and limitations. In section 3 contains the definition of the variants of the h-index along with its advantages and limitations. In section 4, we draw the conclusion of this review work.

\section{h-index}\label{h-index}
To apprise the scientific impact of scholars, several publication based indicators such as the total number of publications, total citation count,  and the average number of citations per paper can be used. However, these indicators have limitations. For example, when considering the total number of publications, a scholar can publish a number of papers, but still have a low scientific attraction in the research community. When considering the total citation count, only a few articles with very high citation counts can hide the very low citation count values of the vast majority of the published articles. When considering the average number of citations per paper, it does not capture  the importance of the high impact articles. Based on the limitations of these indicators, Hirsch \cite{hirsch2005index} proposed a new indicator called h-index. Formally, the h-index is defined as:

\emph{``The h-index of a scholar is h if h of his/her research articles have at least h citation count each and rest of the articles may have h or less citation count."} 

\begin{figure}[h!]
	\centering
	\includegraphics[width=4in,height=3.5in,keepaspectratio]{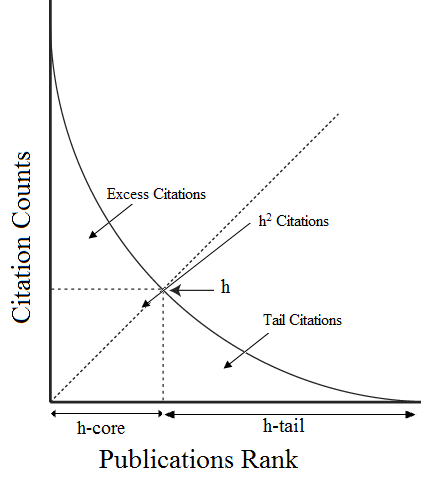}
	\caption{Citation distribution of scholars}
	\label{fig:Citation distribution of scholars}
\end{figure}

The graphical representation of citation distribution of a scholar is shown in Fig.\ref{fig:Citation distribution of scholars}.
   
The graph based definition of h-index is that the h-index is the size of largest square fitted under the curve (i.e. $h^{2}$ citation count), that is the total number of publications which is under the largest square. The area above the largest square fitted under the curve represents the excess citation count of core articles and the area to the right of largest square fitted under the curve represents the total citation count of tail articles, which are not used in h-index computation.

For calculation of h-index, first the publications are arranged in the descending order of their citation count. The publications are then assigned a rank based on their order in the sorted list.  The h-index is the maximum rank a publication where the citation count is equal to or greater than the rank of publication. This index almost covers the productivity and impact of scholars.
\subsection{Advantages and limitations}
The advantage of h-index are the following:
\begin{enumerate}
	\item It is too simple to compute \cite{hirsch2005index} and does not require any data processing \cite{franceschini2010analysis}.
	\item It produces a single number that combines both the quality and the quantity of the scholars' publications \cite{hirsch2005index}. 
	\item It performs better than the total number of articles, total citation count, average citation count, citation per articles, number of highly cited articles, journal citation score \& field citation score in scientific assessment of scholars \cite{hirsch2005index,van2006comparison}.
	\item It can be easily obtained from any publication indexing databases such as Google Scholar, Web of Science etc. \cite{hirsch2005index,alonso2009h,hirsch2014meaning}.
	\item It is closely related to the total number of publications that have significant influence \cite{liu2009properties}.
	\item A small error in the citation distribution does not result in a huge change in the index value \cite{rousseau2007influence,vanclay2007robustness}.  
	\item A single highly cited article does not affect the index value \cite{egghe2009mathematical}. Further, small changes in the citation count of articles do not affect the index value much \cite{hirsch2014meaning}. 
	\item It is also useful in the assessment of the impact of a journal \cite{mingers2012using}.
\end{enumerate}

On the other hand, it suffers with some of the shortcomings, which had been addressed by Hirsch himself and other scholars:
\begin{enumerate}
	\item The citation practice of articles is different in different fields, so it is not useful in comparing the scientific impact of scholars of different fields \cite{hirsch2005index, schreiber2007self, liu2009properties, waltman2011inconsistency,waltman2012inconsistency}. 
	\item It is also not suitable in comparing the scientific impact of scholars' having different research careers\cite{kelly2006h}. 
	\item Once an article is selected for h-core, further citation will not be important in scientific assessment \cite{egghe2006theory,egghe2006improvement}. 
	\item It is very difficult to collect complete citation information of scholars. 
	\item It also considers the self citation count in h-index computation.\cite{van2006comparison}. 
	\item It produces a single natural number, that affects the discriminative power of h-index \cite{tol2008rational, liu2014empirical} when comparing two scholars with the same h-index value. 
	\item The index value never decreases \cite{jin2007r}.
	\item Generally, a research article is  written by a group of scholars who rarely contribute equally. However the h-index gives full credit of citations to all scholars \cite{burrell2007should}. So it is not fair to evaluate the scientific impact of scholars. 
	\item It does not give any extra credit to highly cited articles \cite{van2008generalizing,zhang2009index,Bihari2017}.
	\item It completely ignores the impact of h-tail articles, whereas some of the h-tail articles' citation counts are equal  or very close to the h-index value  \cite{garcia2009multidimensional,Bihari2017}.
	\item The publications inconsistency also does not affect the index value \cite{mahbuba2012diffusion,Bihari2018}.
\end{enumerate}
To overcome the above mentioned limitations, a lot of research has been done by the scholars to provide an efficient alternative to assess the impact of scholars. Some scholars used the properties of h-index and give an effective alternative measure. The alternative measures of h-index are categorized in the following seven categories:
\begin{enumerate}
	\item Complement of h-index (Section \ref{sec:Complement of h-index}).
	\item Based on publication age (Section \ref{sec:Index based on publication age}).
	\item Based on total number of authors (Section \ref{sec:Index based on total number of author}).
	\item Combination of two indices (Section \ref{sec:Indices based on combination of two index}).
	\item Based on excess citation count (Section \ref{sec:Indices based on excess citation count}).
	\item Based on the total number of publications (Section \ref{sec:Index based on total publication}).
	\item Other types of indices (Section \ref{sec:Index based on other variants}).
	\begin{enumerate}
		\item Based on the core tail ratio (Section \ref{sec:Based on core-tail ratio}).
		\item Based on improving h-index to higher values (Section \ref{sec:Based on improvement of h-index to higher value}).
		\item Based on variants of citation process (Section \ref{sec:Based on variants of citation process}).
		\item Miscellaneous indices (Section \ref{sec:Miscellaneous indices}). 
	\end{enumerate}
\end{enumerate}
\subsection{Complement of h-index}\label{sec:Complement of h-index}
As discussed in the above section, the h-index does not give any extra credit to highly cited articles. To overcome this shortcoming of the h-index, several research complemented the h-index and gave alternatives to assess the scientific impact of scholars with the consideration of the impact of the highly cited h-core articles, the total number of h-core articles, the publication age and the total number of publications.  The publication age  is the number of years since the first publication of an author.

The citation count of an article shows the scientific impact of that article and the highly cited articles play an important role in the scientific assessment of scholars. However, the h-index considers only the h number of citation count from the h-core articles, whereas many h-core articles have more than h-citation count. To overcome this big-hit problem of h-index, the \textbf{g-index} has been proposed by \cite{egghe2006theory}. The g-index gives more credit to highly cited articles as well as includes maximum number of articles citation in scientific assessment. Formally, it is defined as:

\emph{``The g-index of a scholar is the largest number g such that the top g articles have at least $g^{2}$ or more citations together."}

The main objective of this index is to overcome the big hit problem of h-index, that is, ``once an article is selected in h-core, further citation is not countable in scientific assessment"  \cite{bornmann2008there}. The main difference between the h-index and the g-index is that the latter is based on the cumulative citation count, whereas the former is based on individual citation count of articles. The small number of highly cited articles give the highest g-index than the greater number of averagely cited papers \cite{alonso2010hg}; this is also one of the main limitations of the g-index \cite{egghe2009econometric}. To overcome the limitations of g-index, \cite{kosmulski2006new} proposed a new index called \emph{h(2)-index}, defined as:

\emph{``The h(2)-index of scholars is the highest natural number such that the h(2) most cited articles received at least $(h(2))^{2}$ citation each."}

For example, if the h(2)-index of a scholar is 10, it means scholar's 10 publications have a minimum of 100 citations each. This index is almost similar to the g-index, the only difference being the method of calculation. The g-index is based on cumulative sum of citation count, whereas the h(2)-index is based on individual citation count. This index considers only the few highly cited articles and penalizes all those scholars who published articles having average citation counts. To overcome the limitations of h(2)-index, \cite{wu2008w} proposed a new variant of h-index named \emph{w-index} and it is defined as:

\emph{``The w-index of scholars is w such that the top w articles have at least 10w citations each."}

For example, if the w-index of a scholar is 5, it means that the scholar's five articles have a minimum of 50 citations each. If the  h(2) index of a scholar is 5, it means scholar's five articles have at least 25 citation each. The w-index is more or less similar to the h(2)-index, but it is less strict than the h(2)-index. It requires only 10 times citation count increment to increase the index value to next. It penalizes all those young scholars who have just started working or those who do not have enough publications. 

\cite{tol2009h} mentioned that the h-index used only highly cited publications, the g-index overcome this shortcoming of the h-index, but it is sensitive to non cited articles. Based on this limitation, the author proposed two different indices called \emph{f-index and t-index} based on harmonic and geometric average. The f-index of an author is defined as:
\begin{equation}
f=\max_{f}\frac{1}{\frac{1}{f}\sum_{p=1}^{f}\frac{1}{cit_{p}}} \geq f
\end{equation}
where $cit_{p}$ is the citation count of $p^{th}$ article. 

The f-index never goes beyond the total number of publications. This index gives higher weight to the least cited articles as compared to the highly cited articles. Based on this limitation, the author proposed a new index called t-index, defined as:
\begin{equation}
t=\max_{t}exp[\frac{1}{t}\sum_{k=1}^{t}ln(cit_{k})]\geq t \leftrightarrow \max_{t}\prod_{k=1}^{t}cit_{k}^{\frac{1}{t}}\geq t
\end{equation}
This index uses many properties of f-index. Both are comparatively difficult to calculate and their values lie between h and g-index. $h \leq f \leq t \leq g$. 

\cite{woeginger2008axiomatic} proposed the other variants of h-index called \emph{w-index}. This is almost similar to h-index. 

\emph{``The w-index of a scholar is the largest value w for which their w articles have at least 1, 2, 3,.....w citation count."}\\
Mathematically, it is defined as:
\begin{equation}
w=\underset{w}{max}(cit_{p}\geq w-p+1), ~~~~~~for~all~p~\leq~w
\end{equation}
where $cit_{p}$ is the citation count of $p^{th}$ article and w is the maximum number of publications. 

The main difference between h-index and the w-index is that the h-index describes the largest square area under the citation curve, whereas the w-index describe the largest isosceles right angle triangle under the citation curve.

The h-index can be used in comparing scientific impact of scholars \cite{xu2015new}. However, the use of h-index  in when comparing two scholars can be controversial, because the total number of core element and the citation count of core elements are often not same for two different scholars having a common h-index value. To overcome this limitation of h-index, \cite{xu2015new}  proposed a new variant of h-index called \emph{Gh-index}. The Gh-index of a scholar $a$, denoted $Gh^{a}$, is defined as:
\begin{equation}
Gh^{a}=\sum_{p=1}^{m}sin g(Cit(Pub_{p}^{a})-GH),~~ where~ sin g(x)= \begin{cases}
1,~~~~ \textrm{$x\geq 0$},\\
0,~~~~ \textrm{x $<$ 0}
\end{cases} 
\end{equation}
where $m$ is the total number of publications of scholar $a$ and $GH$ is the h-index of the scholar.

Above mentioned indices consider only the highly cited articles, however, it seems that the number of highly cited articles and the citation count of the highly cited articles can be different for  scholars with a common h-index value. The h-index fails in comparing the scientific impact of scholars in such a case.  To overcome this limitation, \cite{jin2006h} proposed a new index called \emph{A-index}. Formally, it is defined as: 

\emph{``The A-index of a scholar is the average number of h-core articles citation count."} 

Mathematically, it defined as: 
\begin{equation}
A=\frac{1}{h}\sum_{p=1}^{h}cit_{p}
\end{equation}
where A is the A-index of the scholar,  h represents the h-index and $cit_{p}$ is the citation count of the $p^{th}$ article. 

In the case of high h-index with lower citation count of h-core articles, the A-index may be penalized due to division by h-index \cite{burrell2007h}. In such a case, the one or two highly influential articles reflect the overall index value. So, instead of the average number of citation count, the square root of the sum of h-core articles is more pertinent. Based on this assumption, \cite{jin2007r} proposed another index called \textbf{R-index}, which is further discussed in \cite{glanzel2007r}. Formally, it is defined as:

\emph{``The R-index of scholars is the square root of  the sum of  h-core articles citation count."}

Mathematically, the R-index is defined as:
\begin{equation}
R=\sqrt{\sum_{p=1}^{h} cit_{p}}
\end{equation}
where h represents the h-index and $cit_{p}$ is the citation count of the $p^{th}$ article. 

However, the R-index penalizes all those scholars who have long h-cores. A bigger h-core may penalize the R-index and result in the possibility of getting a lower index value than the scholars having relatively smaller h-cores. To overcome this shortcoming, \cite{panaretos2009assessing} proposed a new index called $R_{m}$-index. Formally, it is defined as:

\emph{``The $R_{m}$-index of scholars is the square root of the sum of square roots of the citation counts of the h-core articles ." }

Mathematically, the $R_{m}$-index is defined as:
\begin{equation}
R_{m}=\sqrt{\sum_{k=1}^{h}cit_{k}^{\frac{1}{2}}}
\end{equation}
where $cit_{k}$ is the citation count of $k^{th}$ article. 

But this index also suffers with the variability in the citation count of the h-core. If h-core articles have less variation, it results in high $R_{m}$-index value. To handle variability in h-core citation count, the coefficient of variation (CV) can be use. Based on CV, the author proposed another indicator called $R_{m-cv}$. The $R_{m-cv}$-index is simply the subtraction of  the CV score of the h-core articles from the $R_{m}$-index. In this regard,\cite{bornmann2007b} mentioned that it is very difficult to set the cutoff value for every author and none of the indicators addressed this issue.  To overcome this limitation, \cite{bornmann2007b} used the field specific reference standard for setting the cutoff value with the help of  ESI (Essential Science Indicator) from Thomson Reuters and proposed a new index called \textbf{b-index}. Formally, it is defined as:

\emph{``The b-index of a scholar is b such that at least b articles belongs to the top 10\% of the publication in a specific field."} 

Mathematically, the b-index is defined as:
\begin{equation}
b=\sum_{y=1}^{Y} \sum_{p=1}^{P}(cit_{py} \geq Pub_{10py})
\end{equation}
where,\\
$b$ is the b-index of scholar, \\$Y$ represents publication year, \\$P$ represents the total publication count, \\$cit_{py}$ represents the citation count of $p^{th}$ article in year $y$, and \\$Pub_{10py}$ is the 10\% article published in year $y$.

The main objective of this index is to identify the field specific prominent actor, because no one is good in all fields. It shows the field specific interest of scholars. But the main limitation of this index is the calculation process. It is difficult to calculate compared to the h-index. 

The average and the square root of the citation count of the h-core articles has been used in scientific assessment of scholars in A-index and R-index respectively. Generally, it seems that the citation distribution of scholars' article is skewed, therefore, instead of the average or the square root of the citation count, the median number of citation is much better to assess the scientific impact of scholars. In this way, \cite{bornmann2008there} proposed a new index called \emph{m-index}. Formally, it is defined as:

\emph{``The m-index of a scholar is the median number of citation count of the h-core articles."}

The $\mu$-index \cite{glanzel2010hirsch} has also been discussed in the same way as the \emph{m-index}. 

\cite{egghe2008h} state that the publication count and the citation count vary from scholar to scholar. The variation between the publication count and the citation count captures the sensitivity of performance of scholars. Based on the sensitivity of the performance changes, they proposed the weighted h-index $(h_{w})$ \cite{egghe2008h}. This index is almost similar to the r-index, but the difference is in the definition of h-core articles. In this index, instead of ranking an article based on its position in the sorted list, the authors used weighted ranking mechanism to rank the articles. The weighted rank of an article is defined as:
\begin{equation}
R_{w}(k) = \frac{\sum_{p=1}^{k}cit_{p}}{h}
\end{equation} 
where $h$ is the h-index and $cit_{p}$ is the citation count of the $p^{th}$ article.

The weighted rank of the $k^{th}$ article is the ratio of cumulative sum of the citation count of the top $k$ articles and the original h-index. Then, the weighted h-index of a scholar is the square root of the sum of the citation counts of the weighted core articles. Mathematically, it is defined as:
\begin{equation}
h_{w}=\sqrt{\sum_{k=1}^{R} cit_{k}}
\end{equation}
where, $cit_{k}$ is the citation count of the $k^{th}$ article and $R$ is the largest rank among all publications such that the $k^{th}$ weighted rank $\leq cit_{k}$.

In research community, the h-index is used in comparing the scientific impact of scholars. However, the h-index does not consider the research career of scholars. Hence the question arises: ``How can we compare the scientific impact of scholars having same h-index but different research careers?" \cite{vaidya2005v} argued that it is not fair to compare a scholar who gives 100\% of his/her time with another scholar who gives only 40\%, though both have the same h-index value. To overcome this limitation the \emph{v-index} was introduced. The v-index of a scholar is the h-index value adjusted with  the publication age. Mathematically, the v-index is defined as:
\begin{equation}
v=\frac{h}{p(y_{this}-y_{0})}
\end{equation} 
where: \\$h$ is the h-index, \\$y_{this}$ is the current year, and \\$y_{0}$ is the year of first publication.

The presumption of h-index is that its value is almost equal to the career length. Based on this concept, \cite{burrell2007hirsch} proposed a new index called \emph{m-quotient} that can be used in comparing scientific impact of scholars with different publication age.

\emph{``The m-quotient of scholars is defined as: m=h/$pub_{age}$."}\\
where, $h$ represents the h-index and  $pub_{age}$ is the publication age. This index is almost similar to the v-index.

As discussed earlier, the h-index can be used in comparing the scientific impact of scholars. If we compare the scientific impact of two different scholars who do not have an equal number of articles, then it is not fair to use h-index as a comparing tool. To overcome this limitation of h-index, several new indicators complimented the h-index in the context of the total number of publications.

\cite{sidiropoulos2007generalized} addressed the above mentioned limitation of h-index and proposed a new index called \emph{Normalized h-index}. Formally, it is defined as:
\begin{equation}
normalized~h-index =\frac{h}{Pub_{count}}
\end{equation}   
where, $h$ represents the h-index and $pub_{count}$ is the total number of articles.

Another similar approach called called \emph{v-index} was proposed by \cite{riikonen2008national}. Formally, it is defined as:

\emph{``The v-index of a scholar is the ratio of the h-index and the total number of publications."}

In recent years, the collaboration among scholars plays an important role in completion of the research work. \cite{hirch2010index} mentioned that an article with 20 citations and published by a group of scholars is played equally in the h-index with other articles with 20 citations but published by single authors. However,  the significance of both articles is not the same. Based on this scenario, the author proposed a new index called $\overline{h}$-index. 

\emph{``The $\overline{h}$-index of scholars is $k$ such that the top k articles have at least k citations each and the co-authors of each article also have an h-index value of at least k."}\\
This index is very difficult to calculate because it requires article citation count as well as co-author's h-index. Further, this index also penalizes articles published with collaborative efforts. Suppose an article got a good number of citations but the co-authors do not have h-index values equal to the citation count, then that article does not contribute to the $\overline{h}$-core. Whereas, an article that has  less citation count, but the co-authors have h-index greater or equal to the citation count, belongs to the $\overline{h}$-core. Another limitation of this index is that it drastically decreases after some time. If an author has collaborated with same age authors, then there is a probability that their $\overline{h}$-index will decrease in future, whereas, had the author collaborated with younger scholars, their $\overline{h}$-index would have increased.

Table \ref{tab:Summary of Complement of h-index} shows the summary of the indices which complimented the h-index.
\begin{table}[h!]
	\vspace{-0.5cm}
	\renewcommand{\arraystretch}{1.3}
	\caption{Summary of Complement of h-index}
	\label{tab:Summary of Complement of h-index}
	\centering
	\begin{tabular}{cp{9.2cm}p{2.0cm}}
		\hline
		\textbf{Index}     &\textbf{Definition}    &\textbf{Publication}\\
		\hline
		g-index & \emph{``The g-index of a scholar is the largest number g such that the top g articles have at least $g^{2}$ or more citations together."} & \cite{egghe2006improvement}\\
		
		h(2)-index &\emph{``The h(2)-index of scholars is the highest natural number such that the h(2) most cited articles received at least $(h(2))^{2}$ citation each."}  &\cite{kosmulski2006new}\\
		
		w-index & \emph{``The w-index of scholars is w such that the top w articles have at least 10w citations each."} &\cite{wu2008w}\\
		
		f-index & $f=\max_{f}\frac{1}{\frac{1}{f}\sum_{p=1}^{f}\frac{1}{cit_{p}}} \geq f$, 	
	where $cit_{p}$ is the citation count of $p^{th}$ article. &\cite{tol2009h}\\
		t-index &
		$t=\max_{t}exp[\frac{1}{t}\sum_{k=1}^{t}ln(cit_{k})]\geq t \iff  \max_{t}\prod_{k=1}^{t}cit_{k}^{\frac{1}{t}}\geq t$ &\cite{tol2009h}\\
		
		Woeginger w-index &\emph{``The w-index of a scholar is the largest value w for which their w articles have at least 1, 2, 3,.....w citation count."} & \cite{woeginger2008axiomatic}\\
		
		Gh-index & 
		$Gh^{a}=\sum_{p=1}^{m}sin g(Cit(Pub_{p}^{a})-GH),~~ where~ sin g(x)= \begin{cases}
		1,~~~~ \textrm{$x\geq 0$},\\
		0,~~~~ \textrm{x $<$ 0}
		\end{cases} $& \cite{xu2015new}\\
		
		A-index	& \emph{``The A-index of a scholar is the average number of h-core articles citation count."}  A= $A=\frac{1}{h}\sum_{p=1}^{h}cit_{p}$,where A is the A-index of the scholar,  h represents the h-index and $cit_{p}$ is the citation count of the $p^{th}$ article. 	& \cite{jin2006h}\\
		
		R-index	& \emph{``The R-index of scholars is the square root of  the sum of  h-core articles citation count."} $R=\sqrt{\sum_{p=1}^{h} cit_{p}}$, where h represents the h-index and $cit_{p}$ is the citation count of the $p^{th}$ article.  &\cite{jin2007r}\\
		
		$R_{m}$-index & \emph{``The $R_{m}$-index of scholars is the square root of the sum of square roots of the citation counts of the h-core articles ." }	$R_{m}=\sqrt{\sum_{k=1}^{h}cit_{k}^{\frac{1}{2}}}$, 
		where $cit_{k}$ is the citation count of $k^{th}$ article. &\cite{panaretos2009assessing}\\
		
		b-index & \emph{``The b-index of a scholar is b such that at least b articles belongs to the top 10\% of the publication in a specific field."} 
		$b=\sum_{y=1}^{Y} \sum_{p=1}^{P}(cit_{py} \geq Pub_{10py})$ & \cite{bornmann2007b}\\
		
		m-index &\emph{``The m-index of a scholar is the median number of citation count of the h-core articles."} &\cite{bornmann2008there}\\
		
		Weighted h-index & \emph{$h_{w}=\sqrt{\sum_{k=1}^{R} cit_{k}}$, where, $cit_{k}$ is the citation count of the $k^{th}$ article and $R$ is the largest rank among all publications such that the $k^{th}$ weighted rank $\leq cit_{k}$. } & \cite{egghe2008h}\\
		
		v-index &  \emph{`` The v-index of a scholar is the h-index value adjusted with the publication age."} $v=\frac{h}{p(y_{this}-y_{0})}$, where h is the h-index, $y_{this}$ is the current year, $y_{0}$ is the first publication year. &\cite{vaidya2005v}\\
		
		m-quotient 	&\emph{``The m-quotient of scholars is defined as: m=h/$pub_{age}$."}
		where, $h$ represents the h-index and  $pub_{age}$ is the publication age.  &\cite{burrell2007hirsch}\\
		
		Normalized h-index & $normalized~h-index =\frac{h}{Pub_{count}}$,  
		where, $h$ represents the h-index and $Pub_{count}$ is the total number of articles. & \cite{sidiropoulos2007generalized}\\
		
		Riikonen v-index &\emph{``The v-index of a scholar is the ratio of the h-index and the total number of publications."}  &\cite{riikonen2008national}\\
		$\overline{h}$-index &\emph{``The $\overline{h}$-index of scholars is $k$ such that the top k articles have at least k citations each and the co-authors of each article also have an h-index value of at least k."} &\cite{hirch2010index}\\	
		\hline 
	\end{tabular}
	\vspace{-0.5cm}
\end{table}

\subsection{Indices based on publication career} \label{sec:Index based on publication age}
Generally, all citation based metrics are solely based on the number of publications and their citation counts. Whenever we compare the productivity of scholars, only  the total number of publications and their citation count is considered. But the productivity of scholars are different at different stages of their careers.  For example, a scholar who is retired or not active in research, but his/her articles are getting regular citations, is considered prominent. Whereas, a young scholar, who has published quite a few papers, but has a smaller number of citations (due to a small career) is not considered as prominent. Based on the above discussion, it is clear that the publication career of the scholars play an important role in the scientific assessment of scholars as well as in comparing the scientific impact of scholars.  

On these lines, \cite{jin2007ar,jin2007r} mentioned that if two scholars have an equal citation count of h-core articles with different publication careers, then the h-index of both scholars is the same. But this is not fair because their publication careers are different. To overcome this limitation of h-index, the AR-index has been proposed, which considers the total career of scholars. Formally,  the AR-index of a scholar is defined as: 
\begin{equation}
AR=\sqrt{\sum_{p=1}^{h} \frac{cit_{p}}{age_{p}}}
\end{equation}
where $cit_{p}$ and $age_{p}$ is the citation count and the age of the $p^{th}$ article respectively.

The main objective of this index is to give equal weightage to all the publications that are either published earlier or recently, and are useful in comparing the scientific impact of scholars having different lengths of publication career. But this index penalizes all those articles that were published earlier. The index value may decrease over time, but helps in estimating the recent scientific impact instead of the total scientific impact. 

Sidiropoulos et. al. (2007)\cite{sidiropoulos2007generalized} proposed the \emph{Contemporary h-index} to give more credit to the citation of newer articles than the older ones. Formally, the Contemporary h-index is defined as:

\emph{``The contemporary h-index of a scholar is $h^{C}$ such that their $h^{C}$ articles have at least score $S^{C} \geq h^{C}$ each."}

The score $S^{C}$ of the $k^{th}$ article is defined as:
\begin{equation}\label{eq:contemporary}
S^{C}_{k}=\gamma (year_{now}-year_{k}+1)^{-\delta} * |cit_{k}|
\end{equation}
where, \\$year_{now}$ is the current year, \\$year_{k}$ is the publication year, \\$cit_{k}$ is the citation count of the $k^{th}$ article, and \\$\gamma,\delta$ are the coefficients set by the user. 

If $\delta$ is 1, then the score of an article is the total citation count divided by the age of the article. It produces very  small values that help to derive a new meaningful variant. The main objective of this index is to give more credit to recent articles rather than the older ones. However, the Contemporary h-index penalizes all those old articles that are continually earning citation till date. To overcome this limitation, \cite{sidiropoulos2007generalized} proposed the \emph{Trend h-index} to measure the current impact of scholars by the recent citation count of all articles. It is defined as: 

\emph{``The trend h-index of a scholar is the largest number $h^{t}$ such that for his/her $h^{t}$ articles, each have a score $S^{t} \geq h^{Ct}$." } 

The score of the $k^{th}$ article is defined as:
\begin{equation}
S^{t}_{k}=\gamma \sum_{\forall c \in cit_{k}}(year_{now}-year_{c}+1)^{-\delta} 
\end{equation}
where the symbols have the same meaning as defined in Eq.\ref{eq:contemporary}.  

This index requires year wise citations of all publications, which is one of the main limitations of  the Trend h-index and makes its computation much more complex than the h-index. Another issue related to this index is the choice of $\gamma$ and $\delta$ parameters. It is very difficult to assign a reasonable value for these parameters.

It is very difficult to differentiate between two scholars having equal h-index as well as equal citation counts in h-core articles.  However, it seems that the h-index of some scholars remains unchanged for some time, while other scholar's citation as well as h-index rises. In this context, the  dynamic h-type-index ($h_{d}$) was designed by \cite{rousseau2008proposal}. This covers the size of core articles and how that size changes with the time.  Formally, it is defined as:
\begin{equation}
h_{d}=R(y).V_{h}(y)
\end{equation}
where,\\
R(y) is the R-index of the scholar at the $y^{th}$ career year, and\\
$V_{h}(y)$ is the recent increment in the h-index for the $y^{th}$ year. 

In this case the value of y is set by the user and it is very difficult to set a reasonable time window for the scientific assessment of a scholar. Instead of considering the total career of a scholar or any fixed time window, a decade based assessment is more precise. Based on this concept \cite{kosmulski2009new} proposed the h-index per decade $(h_{pd})$:

\emph{``The hpd-index of a scholar is hpd such that the hpd articles have at least hpd citation per decade ($ACPD \geq hpd$) each."} 

The adjusted citation per decade is defined as:
\begin{eqnarray}
ACPD=\frac{Citation~from~publication~year~to~year~y}{y-(publication~year-1)}
\end{eqnarray}
where $y$ is the decade year and it must be greater than the publication year. 

Instead of the total time window or decade based scientific assessment,  \cite{fiala2014current} considers only recent three year's citation time window. Pan, R. K. and Fortunato, S. (2014) \cite{pan2014author} and Fortunato, S. (2014) \cite{fortunato2014author} used the last five year citation time window for the scientific assessment of scholars. Instead of fixed time-stamp, a variable time-stamp is much better \cite{schreiber2015restricting}. By using this concept \cite{schreiber2015restricting} considered a variable timestamp $t$, that is either whole publication career or a reasonable time window, and proposed the timed h-index ($h_{t}(y)$). Formally, it is defined as:  

\emph{``The timed h-index of a scholar is the largest integer k such that the k articles have at least k citation count each during the defined citation time window."}\\
The decade year or fixed time window do not consider the overall impact of a scholar. It seems that some of the scholars published articles throughout the career, while some scholars published articles during their PhD career or published occasionally. In both cases, the evaluation of scientific impact of scholars is based on the citation count earned by such articles, but their contribution is very different. In this context, \cite{mahbuba2012diffusion,mahbuba2013year} proposed a set of indicators based on yearly impact of scholars. The year based indices are classified into four categories, where source is year and the items are (i) the total number of publications in a particular year, (ii) the total number of citations earned by all publications published in a particular year, (iii) the total number of citations earned in a particular year from all publications that are published in any year and (iv) diffusion of citation count based on the age of the publications.

\textbf{(i) Career year h-index by publications:} \emph{`` The career year h-index by publications of a scholar is h, if h of his/her publication year has at least h publications each."}\\
To compute the career year h-index by publications, first the total number of publications in each publication year are calculated. Then, they are arranged in descending order of their total publication count. Then, the career year h-index by publications is the maximum rank in which the publication count is equal to or greater than the year-rank. This year based index considers the year wise productivity of scholars. Suppose a scholar productivity is more than others but their scientific impact is very less, then we can not say that the scholar is more prominent than others.

\textbf{(ii) Career year h-index by citations (Item: publication year citation): } \emph{`` The career year h-index by citations of a scholar is h if h of his/her publication years have at least h citation each."}\\
This index considers the total number of citations earned from all articles that are published in a particular year. This shows the productivity year impact of a scholar. Suppose a scholar published more number of articles in their earlier stage of career or selective year, then the index value is very low than the scholars who published articles regularly. So, instead of publications year citation, the citation year citation is more pertinent in scientific assessment.

\textbf{(iii) Career year h-index by citations :} \emph{`` The career year h-index by citations of a scholar is $h$, if $h$ of his/her citation year receives at least sum of $h$ citaton count each."}\\
To compute career year h-index by citations, we first calculate the total number of citations earned in every year from all publications that are published in any year. This index considers the year wise impact of scholars and produces a single number that is equal or less than the total research career. This index may be influenced by the older articles. A good number of earlier published articles affect the index value significantly. Instead of only year wise citation, the age of the publications may also play important role in the scientific assessment. To do this, the diffusion based h-index was designed.

\textbf{(iv) Diffusion based h-index:} \emph{``The diffusion based h-index of scholar is $p$ such that the $p$ year's articles have at least $p$ diffusion citation count each."} 

The diffusion speed of the publication year $Y$ is the sum of the citation counts of all such articles that are published in year $Y$ divided by the age of the publication. All year wise indices consider the year wise impact of scholars rather than the individual publication citation count. It requires year wise publication count, citation count of all articles which are published in respective years. Finally, we can conclude that the all year based can be used as an alternative in scientific assessment of scholars. The career year h-indices use the h-index methodology to compute the overall impact of scholar. As we know that the traditional h-index suffers with big-hit and ignorance of tail-citation issue and the career year h-indices do not account these issue. 

To overcome the big-hit problems or consideration of excess value in scientific assessment of scholars, the year based EM-index has been proposed by \cite{Bihari2018} with three different parameters. The source of the year based EM-index is year and the items are (i) Total publication count in a particular year, (ii) Publication year citation count and (iii) citation year citation. By using these three different item values, the year based EM-index has been designed and named (i) Year based EM-index by publications, (ii) Year based EM-index by publication year citation and (iii) Year based EM-index by citations. The year based EM-index are computed by using the EM-index methodology which is discussed in \cite{Bihari2017}. The year based EM-index produce a set of value along with the global index value. The elements of the year based EM-index help in comparing scientific impact of scholar have similar index value. However, the career year h-indices and the year based EM-index considers only the core item value and leave some important item's value, that may have very near to the core item value. To incorporate the importance of such tail-item value with the consideration of excess citation value, the year based $EM^{'}$-index has been designed by \cite{Bihari2018}. The year based $EM^{'}$-index has been computed by using the methodology of $EM^{'}$-index, which is discussed in \cite{Bihari2017}.  

From the above discussion, it can be concluded that the career of the publication can be used as an important factor in the scientific assessment of scholars and helps in comparing  the impact of junior and senior scientists. The research done till now in comparison of scientific impact of scholars is not sufficient and do not make clarity, hence, need some effort in this context.  Summary of the publication career based index is shown in table \ref{tab:Summary of Index based on age of publication}.
\begin{table}[h]
	\renewcommand{\arraystretch}{1.3}
	\caption{Summary of Index based on age of publication}
	\label{tab:Summary of Index based on age of publication}
	\centering
	\begin{tabular}{cp{9.2cm}p{2.0cm}}
		\hline
		\textbf{Index}     &\textbf{Definition}    &\textbf{Publication}\\
		\hline
		AR-index &\emph{``The AR-index of a scholar is the square root of the sum of the  normalized citations of h-core articles."}   
			$AR=\sqrt{\sum_{p=1}^{h} \frac{cit_{p}}{age_{p}}}$, 
		where $cit_{p}$ and $age_{p}$ is the citation count and the age of the $p^{th}$ article respectively. &\cite{jin2007r}\\
		Contemporary h-index & \emph{``The contemporary h-index of a scholar is $h^{C}$ such that their $h^{C}$ articles have at least score $S^{C} \geq h^{C}$ each."}  & \cite{sidiropoulos2007generalized}\\
		
		Trendy h-index & \emph{``The trend h-index of a scholar is the largest number $h^{t}$ such that for his/her $h^{t}$ articles, each have a score $S^{t} \geq h^{Ct}$." }  &\cite{sidiropoulos2007generalized}\\
		
		Dynamic h-type index & $h_{d}=R(y).V_{h}(y)$, where, R(y) is the R-index of the scholar at the $y^{th}$ career year, and $V_{h}(y)$ is the recent increment in the h-index for the $y^{th}$ year. & \cite{rousseau2008proposal}\\
			\hline
	\end{tabular}
\end{table}
\begin{table}[]
\addtocounter{table}{-1}
\renewcommand{\arraystretch}{1.2}
\footnotesize
\caption{Summary of Index based on total number of author Continue..}
\centering
\begin{tabular}{p{3.0cm}p{9.2cm}p{2.0cm}}
\hline
\textbf{Index}     &\textbf{Definition}    &\textbf{Publication}\\
\hline
		
		hpd-index &\emph{``The hpd-index of a scholar is hpd such that the hpd articles have at least hpd citation per decade ($ACPD \geq hpd$) each ."}  
		$ACPD=\frac{Citation~from~publication~year~to~year~y}{y-(publication~year-1)}$, where $y$ is the decade year and it must be greater than the publication year.  &\cite{kosmulski2009new}\\
		
		Timed h-index &\emph{``The timed h-index of a scholar is the largest integer k such that the k articles have at least k citation count each during the defined citation time window."}\\ &\cite{schreiber2015restricting}\\
		
		Career year h-index by publications & \emph{`` The career year h-index by publications of a scholar is h, if h of his/her publication year has at least h publications each."}  &\cite{mahbuba2012diffusion}\\
		
		Career year h-index by publication year citations & \emph{`` The career year h-index by citations of a scholar is h if h of his/her publication years have at least h citation each."}  &\cite{mahbuba2012diffusion}\\
		
		Career year h-index by citations & \emph{`` The career year h-index by citations of a scholar is $h$, if $h$ of his/her citation year receives at least sum of $h$ citation count each."}  &\cite{mahbuba2012diffusion}\\
		
		Diffusion based h-index &\emph{``The diffusion based h-index of scholar is $p$ such that the $p$ year's articles have at least $p$ diffusion citation count each."} &\cite{mahbuba2012diffusion}\\

		\hline 
	\end{tabular}
\end{table}

\subsection{Indices based on total number of author}\label{sec:Index based on total number of author}
In the research community, most of the research work is done by the group of scholars and the evaluation of the scientific impact of a scholar is based on their articles' citation count. In the scientific assessment process, all authors get full credit of articles citation count. But rarely they contribute equally. To overcome this shortcoming,   \cite{cole1973social} considers only the first author and completely ignore the co-authors. But, it is not fair in the case of multi-authored articles (\cite{wan2007pure}). \cite{lindsey1980production} gives full credit to every scholar. \cite{multiple1981price} used the fractional allocation between all authors, i.e., $1/k$, where $k$ is the total number of scholars. \cite{hodge1981publication}, \cite{sekercioglu2008quantifying} and \cite{hagen2010harmonic} share the credit among all scholars in proportion to their rank.\cite{van1997fractional,trueba2004robust} used the arithmetic counting for credit allocation between author and co-authors. \cite{egghe2000methods} used the geometric series for credit allocation. Several other methods are used to distributed share credit between all scholars \cite{prathap2010there,abbasi2010evaluating,altmann2009evaluating,liu2012modifying,hu2009loads}.

The citation count of an article should be distributed to all co-authors based on their role in the article \cite{batista2006possible}. However, it is very difficult to know the role of each scholar in an article. In this way, simply divide the h-index by the average number of scholars in h-core articles and named the proposed indicator is $h_{i}$-index. It is defined as:

\emph{``The $h_{i}$-index of a scholar is the ratio of the h-index and the average number of scholars in  the h-core articles."} \\
Mathematically, it is defined as:
\begin{equation}
h_{i}=\frac{h}{Avg_{A}}
\end{equation} 
where $h$ is the h-index and $Avg_{A}$ is the average number of scholars from h-core articles. 

Further, \cite{imperial2007usefulness} discusses the impact of $h_{i}$-index. This index penalizes the collaborative effort of scholars because the h-index value is divided by the average number of scholars. If every h-core publication has only one author then the $h_{i}$-index value is equal to h-index. A few high co-authored articles may affect the index value drastically. To overcome this limitation of $h_{i}$-index, \cite{wan2007pure} proposed a new index called \emph{pure h-index}. It is almost similar to the $h_{i}$-index. The only difference is the denominator. In the $h_{i}$-index, the denominator is the average number of scholars in the h-core articles, while in the pure h-index,  the denominator is the square root of the average number of scholars in the h-core articles. The pure h-index is defined as :
\begin{equation}
h_{p}(A)=\frac{h}{\sqrt{E(author)}}
\end{equation}
where $h$ is the h-index and $E(author)$ is the average number of scholars in the h-core articles. 

Further, the authors discussed the equivalent number of co-authors, proportional (arithmetic) and geometric assignment to share credit among all scholars in a multi-authored article.  In case of the equivalent number of co-authors, the credit share of an individual in an article is 1/k (\cite{burrell1995fractional,van1997fractional}), where k is the total number of authors in the article. In case of arithmetic (proportional) assignment (\cite{van1997fractional}), the credit share of an individual(either author or co-author) in an article is defined as :
\begin{equation}
\label{equ:proportional}
C(A,P)=\frac{2(K+1-R)}{K(K+1)}
\end{equation}
Where,\\
$C(A,P)$ is the credit share of scholar A,\\
$K$ is the total number of scholars and\\
$R$ is the rank of the scholar in the $p^{th}$ publication. 

In case of geometric assignment (\cite{egghe2000methods}), the credit share of a scholar in an article is defined as:
\begin{eqnarray}
\label{equ:geometric}
C(A,P)=\frac{2^{(K-R)}}{2^{k}-1}
\end{eqnarray} 
where the symbols have their meaning as defined in Eq. \ref{equ:proportional}

Based on these credit assignment schemes, the pure h-index of a scholar is defined as:
\begin{equation}
h_{p}(A)=\frac{h}{Avg_{Score}}
\end{equation}
where h is the h-index and $Avg_{Score}$ is the average credit share of h-core articles, defined as:
\begin{equation}
Avg_{Score}=\frac{\sum_{P=1}^{h}(C(A,P)}{h}
\end{equation}  
where the symbols have their meaning as defined in Eq. \ref{equ:proportional}.

By using the same credit assignment scheme, the pure R-index of a scholar is defined as:
\begin{equation}
R_{p}(A)=\sqrt{\frac{\sum_{P\in h(pub) }C(A,P)}{E(author)}}
\end{equation}
where,\\
E(author) is the average number of scholar in the h-core articles\\
C(a,p) is defined in Eq. \ref{equ:proportional}.\\
h(pub) is the set of h-core articles.

It seems that a scholar may get high h-index, but he rarely contributes as a core author, while another scholar gets relatively lesser h-index, but mostly contributes as a core author. The h-index and the pure h-index does not account for this issue. To resolve this issue, \cite{chai2008adapted} proposed a new index called adaptive pure h-index ($h_{ap})$. In order to determine the adaptive pure h-index, we first calculate the equivalent numbers of co-authors based on the pure h-index. Then, we compute the effective citation count. The effective citation count is the actual number of citation count divided by the square root of the equivalent number of authors. Then, we rank the publications in the descending order of their effective citation counts. The $h_{ap}$-index lies between $h_{eff}$ and $h_{eff}+1.$, where  $h_{eff}$ is the h-index based on the effective citation count ($cit_{eff}$) . Mathematically, the adaptive pure h-index  ($h_{ap}$) is defined as:
\begin{equation}
h_{ap}=\frac{(h_{eff}+1)* cit_{eff}(h_{eff})-h_{eff} * cit_{eff}(h_{eff}+1)}{cit_{eff}(h_{eff})-cit_{eff}(h_{eff}+1)+1}
\end{equation}

If $h_{eff}$ and $h_{eff}$+1 are equal, then $h_{ap}=cit_{eff}(h_{eff})$. If all articles are single authored, then $h_{ap}$ = h-index.  The adaptive pure h-index is also defined with arithmetic (proportional) and geometric credit assignment schemes.   

Another similar approach the \emph{normalized $h_{i}$-index} was proposed by \cite{wohlin2009new}. It evaluates the scientific impact of scholars using basic h-index with adjusted citation count. It is an extension of $h_{i}$-index. The primary difference between  $h_{i}$-index, pure h-index, adaptive pure h-index and normalized $h_{i}$-index is the distribution of citations among scholars. In this index, the citation distribution of an article to all scholars is  the ratio of total citation count and the total number of scholars ($Cit_{P}/Author\_count(P)$, where $Cit_{P}$ is the citation count  and Author\_count(P) is the total number of scholars of $P^{th}$ article). Then the normalize $h_{i}$-index  of a scholar is the maximum rank in which the adjusted citation count is equal or greater than the rank value.

\emph{``The normalized $h_{i}$-index of a scholar is k, if k of his/her articles have at least k normalized citation count each."}

\cite{egghe2008mathematical} discussed the fractional credit allocation technique to share the credit among scholars in multi-authored articles and proposed the fractional h and g-index. 

\emph{``The fractional h-index ($h_{f}$) of a scholar is $h_{f}$, if $h_{f}$ of his/her articles have at least $h_{f}$ fractional citation count each."}\\
Mathematically, it is defined as:
\begin{equation}\label{equ:fractional}
h_{f}=\underset{K}{max}(K\leq\frac{cit(K)}{Author(K)})
\end{equation}
Where, \\
$h_{f}$ is the fractional h-index of a scholar,\\
cit(K) is the citation count of the $k^{th}$-article, and\\
Author(K) is the total number of authors in $k^{th}$ article. 

The author applied the same technique to the g-index and proposed a new index called fractional g-index.

\emph{``The fractional g-index ($g_{f}$) of scholars is  $g_{f}$, if $g_{f}$ of his/her articles have at least $g_{f}^{2}$ cumulative fractional citation count each."}\\
Mathematically, it is defined as:
\begin{equation}
g_{f}=\underset{p}{max}(\sum_{k=1}^{p}\frac{cit_{k}}{Author(k)} \geq p^{2})
\end{equation}
where the symbols have their meaning as defined in Eq. \ref{equ:fractional} 
Instead of fractional citation count, the fractional ranking of publications is used to design fractional h-index ($h_{F}$) (\cite{egghe2008mathematical}). The effective fractional rank of the $p^{th}$ article is the cumulative sum of the effective rank of successive publications ($R= 1/(Author(p)$). The effective rank of the $n^{th}$ article is defined as:
\begin{equation}
R_{eff}(n)=\sum_{p=1}^{n}\frac{1}{Author(p)}
\end{equation}  
where, $n$ is the total number of successive articles and Author(p) is the total number of authors in the $p^{th}$ article.

Then the fractional h-index is defined as:
\begin{equation}
h_{F} =\underset{R_{eff}(p)}{max}(cit_{p} \geq R_{eff}(p)
\end{equation}
where, $R_{eff}(p)$ is the cumulative effective rank of $p^{th}$ article, $cit_{p}$ is the citation count of $p^{th}$ article. \\
Another similar approach proposed by \cite{schreiber2008share} called $h_{m}$-index. 

Instead of arithmetic and  geometric distribution of citation counts, \cite{hagen2008harmonic} suggested the harmonic counting method. The harmonic share of the $k^{th}$ author in an article is defined as:
\begin{equation}
HC(k)=\frac{(1/k)}{[1+(1/2)+.......+(1/m)]}
\end{equation}
where, k is the rank of author in the $m^{th}$ authored article. 

Based on harmonic credit allocation, the harmonic h-index is defined as:. 

\emph{``The harmonic h-index of a scholar is $H_{h}$, if $H_{h}$ of his/her articles have at least $H_{h}$ harmonic credits each."}

Similar approaches, weighted h-index and weighted citation h-cut, have been  discussed by \cite{abbas2011weighted}. These consider the total number of cited articles and the total number of co-authors. To share the citation credit among scholars, the positionally weighted and the equal weighted mechanisms are used.  In the positionally weighted scheme, the first author gets more credit than the second one  and the second author gets more credit than the third one and so on. Finally the summation of all weights is normalized to  1. The weight of the $k^{th}$ author in the $m^{th}$ authored article is defined as:
\begin{equation}
wt_{k}=\frac{2(m-k+1)}{m(m+1)}
\end{equation} 
In equally weighted scheme all authors get equal weight; i.e. 1/m where m is the total number of authors.

\emph{``The weighted h-index of a scholar is  the largest number k such that their k articles have at least k weighted citation aggregate each."}

Mathematically, the weighted h-index is defined as: 
\begin{equation}\label{equ:weighted h-index}
h_{W}|wt_{k} \in P= h_{W}, if (\min_{k=1}^{h_{W}}\{cit_{k}wt_{k}\} \geq h_{W}, \max_{i \neq k} \{cit_{k}wt_{k}\} \leq h_{W})
\end{equation}
where, \\
P is the total credit score earned by positionally or equally weighted scheme,\\
$cit_{k}$ is the citation count of the $k^{th}$ article and\\
$wt_{k}$ is the weighted score of a scholar in the $k^{th}$ article. 

The weighted citation H-cut of a scholar is the sum of the weighted citation count of the weighted h-core articles.
\begin{equation}
\xi_{w}=\sum_{k=1}^{|h_{w}|} cit_{k}wt_{k}
\end{equation}

where the symbols have their meaning as defined in Eq. \ref{equ:weighted h-index}.

In the last 2 or 3 decades, the number of co-authors has continuously increased \cite{harsanyi1993multiple,kennedy2003multiple,greene2007demise}. This plays a critical role in the distribution of citations among the scholars. If the number of co-authors is more, then the distribution based on the average number of authors, arithmetic counting, geometric counting and the harmonic count is not fair. To resolve this issue, \cite{zhang2009proposal} design a new index called weighted h-index. In weighted h-index, the shared credit of the first and the corresponding authors are 1 and the $j^{th}$ author's share credit is defined as:  
\begin{equation}
WC(j,m)=\frac{2(m-j+1)}{(m+1)(m-2)}, m\geq 4, 2\leq j \leq m-1
\end{equation}
where, m is the total number of authors and j is the rank of the author.\\
For example, let an article published by five authors. Let the first and the last author be the primary and the corresponding author respectively. Then both of the authors get full credit for publication. The rest of the authors earn credit based on their rank using the proportional counting method. By using the weighted citation count, the weighted h-index is defined as:

\emph{``The weighted h-index of a scholar is w, if w of his/her articles have at least w weighted citation count each."}

This index is almost similar to the $h_{i}$-index, the only difference is the credit allocation of the first and the corresponding author. All of the credit allocation schemes consider only the mathematical equation to share the credit among scholars. However, the real scenario is different. The role of every scholar is different in different articles, hence, it is not fair to distribute the citation among scholars by using a mathematical equation. In this way, \cite{hu2010those} categories  the contribution of scholars in three different categories : (i) First author, (ii) Corresponding author and (iii) other author (whose contribution is not defined Based on this, \cite{shapiro1994contributions,hu2009loads}) proposed a new index called $h_{maj}$-index (Majority based index). Based on the contribution of a scholar, the following four different measures can accumulate the performance of scholars: \emph{(i)Overall h-index, (ii)First author h-index, (iii)Corresponding author h-index and (iv)other contributor h-index}.

The overall h-index is the original h-index, the first author h-index is the h-index computed from the citation count of all those articles in which the scholar was present as a first author. The corresponding author h-index is the h-index computed from the citation count of all those articles in which the author is present as a corresponding author. Finally,  the other contributor h-index is the h-index computed form citation count of all those articles in which the scholar is present neither as the main author nor as the corresponding author. The relatively high value of the first author h-index indicates that the author mostly worked as a primary author. The relatively high value of the corresponding author h-index indicates that the author mostly worked as a corresponding author and the relatively high value of the other contributors' h-index shows that the author mostly worked as a supportive author.

Instead of these four types of indicators, \cite{bucur2015updated} categories authors into two categories: the primary author (main author and corresponding author) and non-primary author. With consideration of primary and non-primary author the \emph{Hirsch(p,t)} was proposed.  Formally, it is defined as: 
\begin{equation}
Hirsch(p,t)=(h(p),h(t))
\end{equation} 
where, h(p) represents the h-index computed from the citation count of all those articles in which the author was present as a main or a corresponding author, and h(t) represents the overall h-index. 

Suppose a scholar published 3 articles with 2 citations each. Out of these three, 2 articles were published as the primary author (main and corresponding author). Then the h-index based on these two articles is 2 and the overall h-index is also 2. Hence, the Hirsch(p,t)=2,2.

Another similar approach has been discussed by \cite{wurtz2016stratified}. The author mentioned that the first, second, third and last author's contribution is equal. Based on this scenario,  author proposed a new index called \emph{Stratified h-index}, which combines following four types of indicators: (i) First authorship($h_{1}$), (ii) Second authorship($h_{2}$), (iii) Third authorship($h_{3}$) and (iv) Last authorship h-index ($h_{last}$). The relatively  high value of $h_{1}$ indicates that the author mostly worked as a primary author, the relatively high value of $h_{2}$,$h_{3}$ indicates that the author mostly worked as a secondary author with major contribution, and the relatively high value of  $h_{last}$ indicates that the author mostly work as a senior investigator or supervisor.

Instead of distribution of citation among scholars, Aziz et al. (2013)\cite{aziz2013profit} used the impact of collaborators in the scientific gain of scholars. To do this, the harmonic weighted scheme and the rank of the authors is used to estimate the scientific impact of scholars in an article. The weight of the $k^{th}$ author in $m^{th}$ authored article is defined as:
\begin{equation}
W(k,m)=\frac{1+|m+1-2k|}{1/2~m^{2} +m(1-D)}
\end{equation}
where D is 0, if the article authored by the even number of author and 1/2m, if the article is authored by an odd number of authors.

The sum of the weight of all articles is the number of monograph equivalent. The monograph equivalent is the total number of single authored articles. Then the profit (p)-index of a scholar is defined as:
\begin{equation}
p=1-\frac{ME}{Pub_{all}}
\end{equation}
where, $Pub_{all}$ is the total number of published articles and ME is the monograph equivalent of an author, which is defined as:
\begin{equation}
ME(A)=\sum_{p=1}^{Pub_{all}}W(A,p)
\end{equation}
where, $Pub_{all}$ is the total number of publications and W(A,p) is the weight of the author A in the $p^{th}$ article. 

The value of the profit h-index lies between 0 and 1, where 0 indicates that all articles are written by the primary author, i.e., the contribution of the co-authors is zero, or, the papers are singly authored. 

\cite{prathap2010fractional} proposed \emph{Fractional and Harmonic p-index} to account for the number of authors. In the fractional p-index ($p_{f}$), the fractional credit of an author is the 1/(total number of authors) and is defined as:
\begin{equation}
p_{f}=(Cit_{f}^{2}/P_{f})^{1/3}
\end{equation} 
where $cit_{f}$ is the total fractional citation count and $P_{f}$ is the cumulative sum of fractional rank of co-authors. 

The total fractional citation count of an author is defined as:
\begin{equation}
cit_{f}=\sum_{p=1}^{n}rank_{p}cit_{p}
\end{equation}
where, \\
$n$ is the total number of articles,\\
$rank_{p}$ is the rank of author in the $p^{th}$ article, and\\
$cit_{p}$ is the citation count of the  $p^{th}$ article. 

In harmonic counting method,  the weighted credit of the $k^{th}$ scholar in an $m$ authored article is defined as:
(1/k)/(1+(1/2)+(1/3)+ .......+(1/m)) and the harmonic p-index of a scholar is defined as:
\begin{equation}
p_{h}=(Cit_{h}^{2}/P_{h})^{1/3}
\end{equation}
where, $cit_{h}$ is the total harmonic credit of citation counts and $P_{h}$ is the cumulative sum of the harmonic rank.

Instead of arithmetical allocation of citation counts to an author, \cite{galam2011tailor} used the Tailor Based Allocation (TBA) mechanism  to share credit among scholars and proposed a new index called gh-index.

\emph{``The gh-index of a scholar is k, if k of his/her articles have at least k TBA based fractional citation count each."}

In this mechanism, the extra credit $\alpha $ to the first and the $\beta$ to the last author was given. The tailor based credit allocation of first, last and the $t^{th}$ author is respectively defined as :
\begin{equation}
g(1,n)=\frac{n+\alpha}{T_{n}}
\end{equation} 
\begin{equation}
g(n,n)=\frac{n -1 +\beta}{T_{n}}
\end{equation} 
\begin{equation}
g(t,n)=\frac{n-t}{T_{n}}
\end{equation} 
where, $T_{n}=n(1+n)/2+\alpha +\beta$ and  $n$ is the total number of scholar in an article.

In case an article authored by two authors, the credit allocation has three choices:  two to one third, three to one quarter and one to one half. In case of two to one third, the extra credit is given to the first and the last author is 2 and 1 respectively, in three to one quarter, the extra credit is 1 and 0 respectively,  and in case of one to one half, the extra credit given is 0 and 1 respectively for the first and the last author respectively. In case of articles, authored by more than two authors, the decision of credit allocation depends on choice. 

\cite{liu2011fairly} presented a new mechanism to share credit among corresponding and non-corresponding authors. In this mechanism, the corresponding author gets more credit than the non-corresponding author. The credit of a scholar decreases, when the number of scholar increases. The following steps are used to share credit among scholars:
\begin{enumerate}
	\item First, the sequence of authors is rearranged in the following way: first author, corresponding author and the rest of the authors in the original sequence. For example: an article has been written by four scholars A1, A2, A3 and A4. If A4 is the corresponding author, then the sequence is like A1, A4, A2 and A3.
	\item The credit allocation to the $k^{th}$ author in an m-authored article is defined as:
	\begin{equation}
	p(k,m)= m^{-\frac{1}{t}}k^{-(1-\frac{1}{t})}
	\end{equation}
	where, $t$ is the integral constant greater than one. 
	
	A smaller $t$ value give the credit balance among scholars rather than the maximum $t$. Most Important Authors (MIA), i.e., the first author and the corresponding author, tie for the first rank. Their credit allocation is the average of $p(k,m)$ and is defined as:
	\begin{equation}
	p_{1-rank}(m)=\frac{\sum_{a=1}^{n_{(1-rank)}}m^{-\frac{1}{t}}k^{-(1-\frac{1}{t})}}{n_{(1-rank)}}
	\end{equation} 
	\item The normalized credit score of an individual scholar is defined as:
	\begin{equation}
	p^{'}(k,m)=\frac{p(k,m)}{S_{p}(m)}
	\end{equation}
	and
	\begin{equation}
	p^{'}_{1-rank}(m)=\frac{p_{(1-rank)}(m)}{S_{p}(m)}
	\end{equation}
	where $S_{p}(m)$ is the sum of  all credit given to every author.

	\item The citation allocation of the $k^{th}$ scholar in the article $A$ is defined as:
	\begin{equation}
	CC_{(k,A)}=\begin{cases}
	cit_{A} \times p_{1-rank}(m),~~~~ \textrm{For first and corresponding author}\\
	cit_{A} \times p^{'}(k,m),~~~~ \textrm{For others}
	\end{cases}
	\end{equation}
\end{enumerate}
Based on this credit allocation system two different indices were proposed called CCA h-index ($h_{c}$) and CCA g-index ($g_{c}$)

\emph{``The $h_{c}$-index of a scholar is $h_{c}$, if $h_{c}$ of his/her articles have at least $h_{c}$ allocated citation count each." }

\emph{``The $g_{c}$ index of a scholar is $g_{c}$, if $g_{c}$ of his/her articles have at least $g_{c}^{2}$ citation together."}

Generally, most of the indices give more credit to the first and the corresponding author, however, every author have their own importance in an article. The first and the corresponding author may have done most of the research work and guidance. The other authors may have analyzed the research work \cite{biswal2013absolute}. So we cannot discard the importance of non-primary authors. To consider the importance of non-corresponding authors, the Absolute index(Ab-index) \cite{biswal2013absolute} has been designed. In this index, the first author and the corresponding author get full credit and the rest of the authors get shares in decreasing arithmetic progression.  

\emph{``The Ab-index of a scholar is the sum of the partial credit earned from all articles in which the scholar is present either as a first, corresponding or a co-author."} 

Mathematically, the Ab-index of $k^{th}$ author is defined as:
\begin{equation}
Ab(k)=\sum_{A=1}^{m}PC_{A}(k)
\end{equation} 
where, $PC_{A}(k)$ is the partial credit of scholar $k$ in the article $A$. The partial credit of the first and the corresponding author is equal, and is defined as:
\begin{equation}
PC_{1_{st}/co}(p)=\frac{2*cit_{p}}{k+M}
\end{equation}
where, \\
$PC_{1_{st}/co}(p)$ indicates the  total partial credit of the first or  the corresponding author in the $p^{th}$ article,\\
$cit_{p}$ is the citation count of the $p^{th}$ article,\\
$k$ is the total number of authors and,\\ 
$M$ is the first author or/and the corresponding author.  

In single authored articles, author gets full 100\% credit. In case of one first author, one corresponding author and one other co-author,  the credit share of the first and the corresponding author is 40\% each, and 20\% for the other author. In case of the first author also being the corresponding author, the credit share is 66.67\% and the remaining author gets 33.33\% credit share. In case of $r$ non-primary authors (neither the first nor the corresponding author), then the credit share of $r^{th}$ non-primary author is defined as:
\begin{equation}
PC_{r}=\frac{2*cit_{A}(k-r-CA+1)}{(k+f)(k-f+1)}
\end{equation} 
where, $CA$, $f$, $k$ and $cit_{A}$ is the number of corresponding authors, first author, total number of authors and the citation counts of the article $A$ respectively. 

Clearly, the computation of this index is a lot more complex than the h-index.

\cite{altmann2009evaluating} proposed a new index called RP-index (Research Productivity). This index is based on normalized citation count (each citation count is divided by the age of the publication) and the contribution factor of the individual researcher in the group. It is defined as:
\begin{equation}
NCT_{ak}=\frac{Cit_{k}}{age_{k}}*CF_{ak}, 0 \leq CF_{ak} \leq 1
\end{equation}
where, $NCT_{ak}$ indicates the normalized citation score of author $a$ in article $k$ and $CF_{ak}$ is the contribution factor of author $a$ in article $k$; it lies between 0 \& 1. If the contribution of all authors is equal, then $CF_{ak}$ is 1/(total number of authors). 

\emph{``The RP-index of a scholar is RP, if RP of his/her articles have at least average RP normalized citation count each."}

Mathematically, it is defined as:
\begin{equation}
RP_{k}=max(RP|\frac{1}{RP}\sum_{p=1}^{RP}NCT_{pk}), RP\in P_{all}
\end{equation}
where, $P_{all}$ is the total number of publications of scholar. \cite{altmann2009evaluating} also gives a slightly modified definition of the \textit{RP-index} as:
\begin{equation}
\begin{aligned}
RP_{k}^{t}=\begin{cases} \frac{1}{RP_{k}} \sum_{t=1}^{RP_{k}} NCT_{tk}, &\textrm{$if \frac{1}{RP_{k}} \sum_{t=1}^{RP_{k}} NCT_{tk} < RP_{k}+1 $}\\
RP_{k}+0.99, &\textrm{$if \frac{1}{RP_{k}} \sum_{t=1}^{RP_{k}} NCT_{tk} \geq RP_{k}+1 $}
\end{cases}
\end{aligned}
\end{equation}

The contribution of collaborators also plays an important role in the scientific assessment of a scholar. Based on this concept, \cite{abbasi2009evaluating} and \cite{abbasi2010evaluating} proposed a new index called RC(Researcher Collaboration)-index and the CC\\ (Community Collaboration)-index.

\emph{``The RC-index of a scholar is the largest number R such that their R co-authors have at least R average co-author collaboration value each."}\\
Mathematically, the RC-index is defined as:
\begin{equation}
RC_{k}=max(k|\frac{1}{n}\sum_{j=1}^{n}CCV_{j} \geq n), n\in M
\end{equation} 
where, $M$ is the total number of collaborators and $CCV_{j}$ is the co-author collaboration value of author $k$ with co-author $j$,  and is defined as:
\begin{eqnarray}
CCV_{j}= Total~number~of~collaborations~between ~author(k,j) * RP_{j}^{'}
\end{eqnarray}
where $RP_{j}^{'}$ is the RP-index of scholar j. 

After a slight modification of the RC-index, the $RC^{'}$-index is defined as:
\begin{equation}
\left.\begin{aligned}
RC^{'}_{k}=\begin{cases} \frac{1}{RC_{k}} \sum_{j=1}^{RC_{j}}CCV_{c}, & \textrm{$if \frac{1}{RC_{k}} \sum_{j=1}^{RC_{j}}CCV_{c} < RC_{j}+1$}\\
RC_{k}+0.99,  & \textrm{$if \frac{1}{RC_{k}} \sum_{j=1}^{RC_{j}}CCV_{c} \geq RC_{j}+1$}.
\end{cases}
\end{aligned}\right.
\end{equation} 
The $RC^{'}_{k}$ values lies between $RC_{k}$ and $RC_{k}$ +1.

The Community collaboration index called \emph{CC-index} is defined as:
\begin{equation}
CC_{k}=max(k|\frac{1}{n}\sum_{j=1}^{n}RC_{j} \geq n), n\in M
\end{equation} 
After a slight modification of the CC-index, the $CC^{'}$-index is def]ined as:
\begin{eqnarray}
CC^{'}_{k}=\begin{cases} \frac{1}{CC_{k}} \sum_{j=1}^{CC_{j}}RC_{c}, & \textrm{$if \frac{1}{CC_{k}} \sum_{j=1}^{CC_{j}}RC_{c} < CC_{j}+1$}\\
CC_{k}+0.99,  & \textrm{$if \frac{1}{CC_{k}} \sum_{j=1}^{CC_{j}}RC_{c} \geq CC_{j}+1$}.
\end{cases}
\end{eqnarray}

In this section, we have discussed a number of indices. They consider the number of co-authors in the scientific assessment of scholars and the research articles. Several indices consider the total number of authors of h-core articles. Some of them are defined through somewhat complex mathematical equations, some consider the impact of only the first and the corresponding author, and so on. These all are only the assumption because a research article does not have any information related to author contributions. After a long journey of scientific assessment of scholars, the sharing of credits among scholars is still an open challenge. Summary of the indices based on total number of author shown in table \ref{tab:Summary of Index based on total number of author}
\begin{table}[h!]
	\renewcommand{\arraystretch}{1.3}
	\caption{Summary of Index based on total number of author}
	\label{tab:Summary of Index based on total number of author}
	\centering
	\begin{tabular}{cp{9.2cm}p{2.0cm}}
		\hline
		\textbf{Index}     &\textbf{Definition}    &\textbf{Publication}\\
		\hline
		$h_{i}$-index &\emph{``The $h_{i}$-index of a scholar is the ratio of the h-index and the average number of scholars in  the h-core articles."} $h_{i}=\frac{h}{Avg_{A}}$, where $h$ is the h-index and $Avg_{A}$ is the average number of scholars from h-core articles.  &\cite{batista2006possible}\\
		
		Pure h-index & \emph{$h_{p}(A)=\frac{K}{\sqrt{E(author)}}$, where $h$ is the h-index and $E(author)$ is the average number of scholars in the h-core articles.} &\cite{wan2007pure}\\
		
		Pure R-index & \emph{$R_{p}(A)=\sqrt{\frac{\sum_{P\in h(pub) }C(A,P)}{E(author)}}$}& \cite{wan2007pure}\\
		
		Adaptive pure h-index($h_{ap}$) & \emph{$h_{ap}=\frac{(h_{eff}+1)* cit_{eff}(h_{eff})-h_{eff} * cit_{eff}(h_{eff}+1)}{cit_{eff}(h_{eff})-cit_{eff}(h_{eff}+1)+1}$, where $h_{eff}$ represents h-index based on effective citation count.} &\cite{chai2008adapted}\\
	
		Normalized $h_{i}$-index &\emph{``The normalized $h_{i}$-index of a scholar is k, if k of his/her articles have at least k normalized citation count each."} &\cite{wohlin2009new}\\

		Fractional h-index ($h_{f}$) & \emph{``The fractional h-index ($h_{f}$) of a scholar is $h_{f}$, if $h_{f}$ of his/her articles have at least $h_{f}$ fractional citation count each."} &\cite{egghe2008mathematical}\\
		
		Fractional g-index ($g_{f}$) & \emph{The fractional g-index ($g_{f}$) of an author is  $g_{f}$ such that the top $g_{f}$ papers has at least $g_{f}^{2}$ cumulative fractional citation count each.} &\cite{egghe2008mathematical}\\
		
		Harmonic h-index &\emph{``The harmonic h-index of a scholar is $H_{h}$, if $H_{h}$ of his/her articles have at least $H_{h}$ harmonic credits each."} &\cite{hagen2008harmonic}\\	
		
		Weighted h-index &\emph{``The weighted h-index of a scholar is  the largest number k such that their k articles have at least k weighted citation aggregate each."} &\cite{abbas2011weighted}\\
		
		Zhang Weighted h-index &\emph{``The weighted h-index of a scholar is w, if w of his/her articles have at least w weighted citation count each."} &\cite{zhang2009proposal}\\
		
		Hirsch(p,t)	&\emph{The Hirsch(p,t) is the combination of sum of h-index of author based on citation count of all those article where author present as a main \& corresponding author and overall h-index of an author.} (Hirsch(p,t)=h(p),h(t)) & \cite{bucur2015updated}\\
		
		Profit h-index ($P_{h}$) & $p_{h}=1-\frac{h_{a}}{h}$, Where $h_{a}$ is the adjusted h-index of author and h is the actual h-index of the corresponding author. &\cite{aziz2013profit}\\
		
		Fraction p-index ($p_{f}$) & $p_{f}=(Cit_{f}^{2}/P_{f})^{1/3}$, where $cit_{f}=\sum_{p=1}^{n}rank_{p}cit_{p}$ &\cite{prathap2010fractional}\\
		
		Harmonic p-index($p_{h}$) &$p_{h}=(Cit_{h}^{2}/P_{h})^{1/3}$ &\cite{prathap2010fractional}\\
		
		gh-index &\emph{``The gh-index of a scholar is k, if k of his/her articles have at least k TBA based fractional citation count each."} &\cite{galam2011tailor}\\
		
		$h_{c}$-index  &\emph{``The $h_{c}$-index of a scholar is $h_{c}$, if $h_{c}$ of his/her articles have at least $h_{c}$ allocated citation count each." } &\cite{liu2011fairly}\\
		
		$g_{c}$ index	&\emph{``The $g_{c}$ index of a scholar is $g_{c}$, if $g_{c}$ of his/her articles have at least $g_{c}^{2}$ citation together."} &\cite{liu2011fairly}\\
		
		Ab-index 	& \emph{``The Ab-index of a scholar is the sum of the partial credit earned from all articles in which the scholar is present either as a first, corresponding or a co-author."} ($Ab(k)=\sum_{A=1}^{m}PC_{A}(k)$,  where $PC_{A}(k)$ is the partial credit of author k in article A.) &\cite{biswal2013absolute}\\
		
		RP-index (Research Productivity) &\emph{``The RP-index of a scholar is RP, if RP of his/her articles have at least average RP normalized citation count each."} &\cite{altmann2009evaluating}\\
		
		RC-index	&\emph{``The RC-index of a scholar is the largest number R such that their R co-authors have at least R average co-author collaboration value each."}&\cite{abbasi2009evaluating}\\
		
		CC-index &\emph{$CC_{k}=max(k|\frac{1}{n}\sum_{j=1}^{n}RC_{j} \geq n), n\in M$} &\cite{abbasi2010evaluating}\\ \hline
	\end{tabular}
\end{table}

\subsection{Indices based on combination of two indices}\label{sec:Indices based on combination of two index}
To measure the scientific impact of scholars, several indices were proposed and every index measures the impact with different parameters. In order to find the global impact of a scholar, several parameters should be addressed that cover at least total productivity and the impact of scholars. Any single index covers only one parameter, so the combination of advantages of two or more indices is more precise \cite{martin1996use,van2003holy}. In this context, the hg-index, $q^{2}$-index and $h_{mc}$-index were proposed to assess the scientific impact of scholars. 

The h-index considers only the highly cited articles and leaves a number of articles, however, the g-index tries to use a maximum number of articles\cite{alonso2010hg}.  But both of the indices do not meet the sufficient requirement of the scientific assessment of scholars. So, the combination of these two indices called \emph{hg-index} could be an effective alternative to assess the scientific impact of scholars. Formally, it is defined as:

\emph{``The hg-index of a scholar is the geometric mean of the h and g-index."} 

Mathematically, the hg-index is defined as:
\begin{equation}
hg=\sqrt{h.g}
\end{equation}
where, h and g represents the h and the g-index value of scholar.\\
The hg-index combines the advantages of h-index and g-index to produce a value near to the h-index than to g-index. The lower h-index penalizes the g-index and produces a lower index value. 

\cite{cabrerizo2010q2} categorizes the indices in two categories: the  first one is based on  the productive core of a scholar like h-index(\cite{hirsch2005index}), g-index(\cite{egghe2006theory}), hg-index(\cite{alonso2010hg}) \& $h^{(2)}$ -index(\cite{kosmulski2006new}), while the second one is  based on the impact of core papers like a-index (\cite{jin2006h}), m-index(\cite{bornmann2008there}), AR-index(\cite{jin2007r}) \& $h_{w}$-index(\cite{egghe2008h}). But one measure alone, either based on the productive core or the impact of the productive core,  is not suitable for the scientific assessment of scholars. To overcome this limitation of scientific assessment process, \cite{cabrerizo2010q2} proposed a new index called \emph{$q^{2}$-index} that considers both the number of productive core articles  and the impact of core articles. Formally,   

\emph{``The $q^{2}$-index of scholars is the geometric mean of h-index and m-index."}

Mathematically, it is defined as:
\begin{equation}
q^{2}=\sqrt{h.m}
\end{equation}
where, the h and m is the h and m-index value of scholar.\\
The $q^{2}$-index of a scholar is nearer to the h-index than the m-index. The low value of h-index penalizes the m-index and the resultant $q^{2}$-index is relatively low. 

Another index designed to combine the properties of $h_{m}$-index (\cite{hagen2008harmonic}) and the $h_{c}$-index (\cite{liu2011fairly}) is the $h_{mc}$-index (\cite{liu2012modifying}). In $h_{m}$-index, the citation count is distributed equally to all the scholars ($cit/n$, where $cit$ is the citation count and $n$ is the total number of scholars). Instead of this, the author used combined credit allocation for most important author (MIA) (\cite{liu2011fairly} and \cite{schreiber2008share}) and the cumulative sum of the combined credit allocation, named effective paper count, is used  for publication ranking. Mathematically, it is defined as:
\begin{equation}
p_{eff}(k)=\sum_{p=1}^{k}p(rk(p),m(p)
\end{equation} 
where rk(p) is the rank of author in article $p$ and $m(p)$ is the  number of authors in paper $p$. 

\emph{``The $h_{mc}$-index of a scholar is k, if k of his/her articles have at least k effective article count each."}\\
Mathematically, it is defined as:
\begin{equation}
h_{mc}=\max_{k}(p_{eff}(k) \leq cit(k)
\end{equation}
where, $p_{eff}(k)$ is the effective paper count of the $k^{th}$ article and $cit(k)$ is the citation count of the $k^{th}$ article. If the tail articles earn extra citation count in the near future  and core articles citations remain unchanged, then $h_{mc}$-index will increase. 

In this section, we have discussed all those indices that combine the advantages of two indices. From the above discussion, we can conclude that the combination of two indices could be an effective alternative to citation based indices, but still, each has limitations to emerge as a single index to completely asses the scientific impact of scholars. The summary of index based on the combination of two indices shown in table \ref{tab:Summary of Index based on combinaiton of two index}.
\begin{table}[!h]
	\renewcommand{\arraystretch}{1.3}
	\caption{Summary of Index based on combination of two index}
	\label{tab:Summary of Index based on combinaiton of two index}
	\centering
	\begin{tabular}{cp{9.2cm}p{2.0cm}}
		\hline
		\textbf{Index}     &\textbf{Definition}    &\textbf{Publication}\\
		\hline
		hg-index &\emph{``The hg-index of a scholar is the geometric mean of the h and g-index."}  ($hg=\sqrt{h.g}$) &\cite{alonso2010hg}\\
		
		$q^{2}$-index &\emph{``The $q^{2}$-index of scholars is the geometric mean of h-index and m-index."} ($q^{2}=\sqrt{h.m}$) &\cite{cabrerizo2010q2}\\
		
		$h_{mc}$-index	&\emph{``The $h_{mc}$-index of a scholar is k, if k of his/her articles have at least k effective article count each."} ($h_{mc}=\max_{k}(p_{eff}(k) \leq cit(k)$) &\cite{liu2012modifying}\\
		\hline 
	\end{tabular}
\end{table}

\subsection{Indices based on excess citation count}\label{sec:Indices based on excess citation count}
The scientific impact of an article is measured in terms of citation count and the scientific impact of a scholar is computed based on the citation count of articles. An article that earned more citations than others is considered more influential and a scholar having a higher index value than others is considered a prominent actor. At some stage, further addition in citation counts of an article does not contribute in improving the values of most of the popular indices, but every citation count has its own importance in the scientific assessment of scholars. Such citations are called excess citations. To account for the importance of excess citation count, several indices were designed.

Let scholar $X$ publish 10 articles with 100 citations each and scholar $Y$ also publish 10 articles, but with 10 citations each. In both cases, the h-index value is equal, however, the scientific impact of scholar $X$ is more than scholar $Y$. In another case, let scholar $X$ publish five articles with 100 citations each and  scholar $Y$ also publish 10 articles, but with 8 citations each. The h-index of scholar $Y$ is more than scholar $X$, but the total scientific impact of scholar $Y$ is significantly less than that of scholar $X$. To overcome this limitation of h-index, \cite{zhang2009index} proposed a new index called \emph{e-index} which considers only the excess citation. The e-index of a scholar is defined as:
\begin{equation}
e^{2}=\sum_{p=1}^{h}(cit_{p}-h) = \sum_{p=1}^{h} cit_{p}-h^{2}
\end{equation}
where, $cit_{p}$ is the  citation count of the $p^{th}$ article and $h$ represents the h-index. This index considers only the excess number of citation counts and completely ignore the core citations. This is one of the main limitations of e-index. 

The \emph{j-index} proposed by \cite{todeschini2011j} uses the weighted citation incremental approach. Here, the author suggested 12 arbitrary weighted citation increments ($\Delta$wc) like 500, 250, 100, 50, 25, 10, 5, 4, 3, 2, 1.5 \& 1.25 and the corresponding weight($wt$) as $1/R$, where $R$ is the rank  of weighted citation increment. Formally, the j-index is defined as:
\begin{equation}
j=h+\frac{\sum_{r=1}^{12} wt_{r}M_{r}(h \Delta h_{r})}{\sum_{r=1}^{12}wt_{r}}
\end{equation} 
where, $M_{r}(h \Delta h_{r})$ is number of publications that have at least $h \Delta h_{r}$ citations each.

From the above discussion, we can conclude that neither the e-index nor the j-index fairly account the excess citation count in the scientific assessment of scholars. The e-index considers only the excess citation count of h-core articles and completely ignores the core citation count. Hence, it would be difficult to imagine the scientific assessment of scholars without the core citation count. The j-index works on the concept of weighted incremental approach and how the weights are incremented is one much debatable. In this context one recent index has been proposed by \cite{Bihari2017} named \emph{EM-index}. The EM-index of an author is the square root of sum of the component of EM-index. The component of the EM-index is the h-index calculated at multi-level. The first component of EM-index is the original h-index, second component is the h-index from the excess citation count of h-core article. Third component is the h-index from second level excess citation count and so on.  Mathematically, the EM-index is defined as:
\begin{equation}
EM=\sqrt{\sum_{c=1}^{m}E_{c}}
\end{equation}
where m is the total number of component of EM-index. This index helps us to differentiate two different researcher with same h-index.  
 The summary of index based on the combination of two indices shown in the table \ref{tab:Summary of Index based on excess citation count}.
\begin{table}[!h]
	\renewcommand{\arraystretch}{1.3}
	\caption{Summary of Index based on excess citation count}
	\label{tab:Summary of Index based on excess citation count}
	\centering
	\begin{tabular}{cp{9.2cm}p{2.0cm}}
		\hline
		\textbf{Index}     &\textbf{Definition}    &\textbf{Publication}\\
		\hline
		e-index &$e^{2}=\sum_{p=1}^{h}(cit_{p}-h) = \sum_{p=1}^{h} cit_{p}-h^{2}$ &\cite{zhang2009index}\\
		j-index &$j=h+\frac{\sum_{r=1}^{12} wt_{r}M_{r}(h \Delta h_{r})}{\sum_{r=1}^{12}wt_{r}}$ &\cite{todeschini2011j}\\
		EM-index & $EM=\sqrt{\sum_{c=1}^{m}E_{c}}$ & \cite{Bihari2017}\\
		\hline 
	\end{tabular}
\end{table}
\subsection{Indices based on total publication}\label{sec:Index based on total publication}
The scientific product of a scholar is research articles and every article has its own impact on the research career of a scholar. However, the scientific assessment of a scholar is done with the help of only a few numbers of high impact articles and leave many articles that have citation counts near to the h-index value. It seems that the articles with citation count equal to or a little less than the h-index value are not used in the scientific assessment of a scholar.  Such articles, referred to as h-tail articles, are obviously important, however, even the h-tail articles with lesser citations can be significant in assessing the impact of a scholar, and hence, should not be ignored.  Motivated by this, several studies have been done to incorporate the impact of all those articles that are cited at least once.

\begin{figure}[!h]
	\centering
	\includegraphics[width=4in,height=2.35in,keepaspectratio]{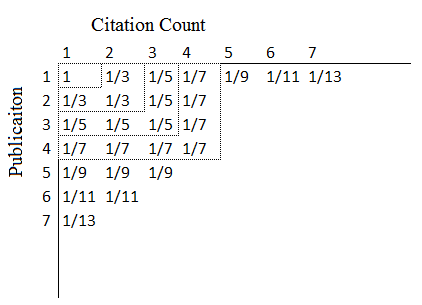}
	\caption{Score of article in tapered h-index}
	\label{Score of article in tapered h-index}
\end{figure}

\cite{anderson2008beyond} proposed a new index called \emph{tapered h-index} based on the total number of cited articles. Tapered h-index is defined based on the Ferrers graph of publications and their citation counts. This graph assigns positive scores to all publications with non-zero citation count. To score an h-index value of 1, only one article with one citation count is required, to score a value of 2, a scholar requires at least 2 publications with 2 citations each. The increment of h-index value from 1 to 2  requires 3 more citations: 1 for the existing article and 2 for the new article. This increment is shown in the Ferrer graph with the contribution score of 1/3 in Fig. \ref{Score of article in tapered h-index}. The figure also shows the score of all other articles.
In the Ferrer graph, a column  represents a partition of citation counts of the articles and a row represents the total citation count. The largest square in the Ferrer graph is called Durfee square; its length represents the h-index value. The score of any citation in the Ferrer graph is defined as:$1/(2l-1)$, where l is the length of any side in the Ferrer graph. The score of an article is simply the sum of the score at every level in the Ferrer graph. The score of the  $p^{th}$ article in the Ferrer graph is defined as:
\begin{numcases}
{h_{T}(p)=}
\frac{cit_{p}}{2p-1},  & if $cit_{p} \leq p$,
\\
\frac{p}{2p-1}+\sum_{p+1}^{cit_{p}}\frac{1}{2p-1}, & if $cit_{p} > p$.
\end{numcases}
where, $cit_{p}$ is the citation count of the $p^{th}$ article. Finally, the tapered h-index ($h_{T}$) of a scholar is the sum of the $h_{T}$ scores of all cited articles. Formally, it is defined as:
\begin{equation}
h_{T}=\sum_{p=1}^{m}h_{T}(p)
\end{equation} 
This index tries to incorporate all citation count in the scientific assessment of scholar, but the computation process of this index is little bit complex. 

In this way, \cite{garcia2009multidimensional} proposed the \emph{multidimensional h-index} to account for the importance of all cited articles.  To compute the multidimensional h-index, first the citation count of all those articles which has at least one citation count is stored in an m-dimensional vector in descending order.  Then, the multidimensional h-index is defined as: H=($h_{1}$, $h_{2}$, $h_{3}$, ......, $h_{p}$), where $h_{1}$ is the first h-index from all $m$ publications, $h_{2}$ is the second subsequent h-index from  first tail (m-$h_{1}$) publications, and so on, until the entire publication citations are exhausted. The component of multidimensional h-index should be such that $h_{1}~\geq ~ h_{2}~ \geq h_{3}$.... .  The $k^{th}$ component of the multidimensional h-index is defined as:
\begin{equation}
h_{k}=p-\sum_{m=0}^{k-1}h_{m}
\end{equation}
where, $p$ is the largest integer in the range $\sum_{m=0}^{k-1}h_{m} + 1 \leq p \leq k $ that satisfies $cit_{p}\geq p-\sum_{m=0}^{k-1} h_{m}$ (where $h_{0}=0$). 

The total number of components in multidimensional h-index is always less than or equal to the total cited publication count.  If all articles have a single citation, then the total number of components in the multidimensional h-index is equal to the total number of publications, otherwise, it is always less than the total cited publication count. The number of components and their values vary from scholar to scholar. This shows the significant difference between scholars. 

For example, let a scholar publish 20 articles. Out of these 20, let 17 articles have at least one citation count: Cit=(100, 20, 20, 17, 10, 10, 9, 9, 8, 8, 5,  4, 4, 3, 2, 2, 1). Then the components of the multi-dimensional h-index are \{8, 4, 2, 2, 1\}. If we take the reasonable minimum component value to be 4, then the components of the multi-dimensional h-index are \{8, 4\}. The sum of the component of the multidimensional h-index is 17 equal to the  total number of cited publications. But this index does not produce a global index value to show the overall impact of scholars. To overcome this limitation, \cite{todeschini2016handbook} defined a new index called \textit{iteratively weighted h-index}. The iteratively weighted h-index of a scholar is defined as:
\begin{equation}
iw(h)=\sum_{c=1}^{p}\frac{h_{c}}{c}
\end{equation}
where, $p$ is the total number of components and $h_{c}$ is the $c^{th}$ component of the multidimensional h-index.

 \cite{franceschini2010analysis} proposed a new variant with additional sign to account all the cited publications in scientific assessment. The additional sign is the total time-range of all those articles that have at least one citation count. It is defined as:
\begin{equation}
f=range_{k \in p_{all} }(yr_{1}, yr_{2}, yr_{3},........,yr_{k})+1
\end{equation} 
where, $P_{all}$ is the set of all cited publications and $yr_{k}$ is the publication year of $k^{th}$ article.

The main feature of this indicator is that it never decreases and is never more than the total publication career. For example, to compute the f-range, consider an artificial example of the publication history of scholar $A$ as given in Table \ref{tab:Example  of f-value}. Here, $A$ publishes 10 articles such that out of these 10, 8 articles have at least 1 citation count.
\begin{table}[h!]
	\renewcommand{\arraystretch}{1.3}
	\centering
	\caption{Example  of f-value}
	\label{tab:Example  of f-value}
	\begin{tabular}{p{1.5cm}cccccccccc}
		\hline
		\textbf{Year}     & 2000	  &2000	&1989	&2001	&	2005	&2009	&2010	&2010	&1990	&2014	\\
		\textbf{Citation Count}&40	&30	&29		&12		&5	&5 	&5			&1	&0 &0\\
		\textbf{Rank}	&1	&2	&3	&4	&5	&6	&7 		&8	&9 &10	\\\hline
	\end{tabular}
\end{table}

The h-index of scholar is 5  and the f-value of scholar A = \{2000, 1989, 2001, 2005, 2009, 2010\} + 1 = 7. Here, +1 is done to account for the preparation of the first article. The scientific impact of a scholar with this additional sign is categorized in the following four categories:
\begin{enumerate}
	\item Low h-index and low f-value: It indicates new scholars.
	\item Low h-index and high f-value: It indicates the scholar working from a long time but their research impact is relatively less.
	\item High h-index and low f-value: It indicates a young prominent scholar. 
	\item High h-index and high f-value: It indicates the scholar worked hard in the research community from long time.
\end{enumerate}
This index helps in comparing the scientific impact of scholars having different research careers. 

\cite{smith2015platinum} proposed the \emph{Platinum H}-index  that covers the total citation count, total research career and the total publication count. Formally, it is defined as: 
\begin{equation}
Platinum H=\frac{H}{CL} \times \frac{Cit_{all}}{Pub_{count}}
\end{equation}
where, $H$ is the h-index, $CL$ is the career length, $Cit_{all}$ is  the total citation count and $pub_{count}$ is the publication count. The computation process of the platinum H-index is very simple and comparable to that of other than the h-index. The Platinum H-index  considers the total number of articles; these articles can be either cited or otherwise. However, the logic behind inclusion of non-cited publications in scientific assessment is not clear. Motivated by this, \cite{yaminfirooz2015multiple} proposed  a new index called multiple h-index (mh-index). 

\emph{``The mh-index of a scholar is the square root of the product of h-index at every level and the square of citation count of an article is divided by the age of the publication."} 

Mathematically, it is defined as:
\begin{equation}
mh=\sqrt{\sum_{p=1}^{n}}\sum_{k=1}^{h_{p}}\frac{h_{p}cit_{k}^{2}}{age_{k}}
\end{equation}
where,\\
$h_{p}$ is  the scholar's $p^{th}$ h-index,\\
$cit_{k}$ is the citation count of  $k^{th}$ article, and\\
$age_{k}$ is the age of $k^{th}$ article. 

To calculate the mh-index, first calculate the h-index from all those articles that have at exactly one citation count. Then calculate the h-index from all those articles that have 2 citation count,  and so on. Finally, calculate the mh-index. 

All of the indices which considers the total number publications do not account the importance of citation count of all articles. To overcome this limitations \cite{Bihari2017} proposed a new index called $EM^{'}$-index. The $EM^{'}$-index of an author is the square root of sum of the component of $EM^{'}$-index. The component of the $EM^{'}$-index is the h-index, calculated at multi-level. The first component of $EM^{'}$-index is the original h-index, after that subtract the h-index value from core-publications citation count, then the second component of $EM^{'}$-index is h-index value from updated citation data set. This process will continue until the all citation count exhausted or the all publications citation remain one or only one article citation count is more than zero. Mathematically, the $EM^{'}$-index is defined as:
\begin{equation}
EM^{'}=\sqrt{\sum_{c=1}^{m}E_{c}^{'}}
\end{equation} 
where m is the total number of component in $EM^{'}$-index.

In this section, we have discussed a number of indices that consider the total number of publications that have at least one citation count in the scientific assessment of scholars. Every indicator gives a  perspective to consider the impact of the total number of articles, but none of the indices consider the total number of scholars and the total citation counts with the total number of publications. This may provide a good alternative in the scientific assessment of scholars.
The summary of index based on the total publication count shown in the table \ref{tab:Summary of Index based on total publication count}.
\begin{table}[!h]
	\renewcommand{\arraystretch}{1.3}
	\caption{Summary of Index based on total publication count}
	\label{tab:Summary of Index based on total publication count}
	\centering
	\begin{tabular}{cp{9.2cm}p{2.0cm}}
		\hline
		\textbf{Index}     &\textbf{Definition}    &\textbf{Publication}\\
		\hline
		Multidimensional h-index &\emph{Multidimensional h-index is the set of h-index. It captures all publication which has at least one citation count. Multidimensional h-index is defined as: H=($h_{1}$, $h_{2}$, $h_{3}$, ......, $h_{p}$). Where $h_{1}$ is first h-index, $h_{2}$ is subsequent h-index of k-$h_{1}$ papers and so on.} &\cite{garcia2009multidimensional}\\
		
		Iteratively weighted h-index & $iw(h)=\sum_{c=1}^{p}\frac{h_{c}}{c}$, where, $p$ is the total number of components and $h_{c}$ is the $c^{th}$ component of the multidimensional h-index. & \cite{todeschini2016handbook}\\
		
		Platinum H-index &$Platinum H=\frac{H}{CL} \times \frac{Cit_{all}}{Pub_{count}}$, where, $H$ is the h-index, $CL$ is the career length, $Cit_{all}$ is  the total citation count and $pub_{count}$ is the publication count. &\cite{smith2015platinum}\\
		
		mh-index & \emph{``The mh-index of a scholar is the square root of the product of h-index at every level and the square of citation count of an article is divided by the age of the publication."} &\cite{yaminfirooz2015multiple}\\
		
		$EM^{'}-index$ & $EM^{'}=\sqrt{\sum_{c=1}^{m}E_{c}^{'}}$ & \cite{Bihari2017}\\
			\hline 
	\end{tabular}
\end{table} 

\subsection{Indices based on other variants}\label{sec:Index based on other variants}
Instead of the total number of publications, the age of publications, the number of co-authors, the excess citation count, etc., several indices consider other parameters such as a successful paper, the total number of citers (reader), the citation curve and many more to assess the scientific impact of scholars.

\subsubsection{Based on core \& tail citation} \label{sec:Based on core-tail ratio}
Generally, the citation count of a scholar's articles is categorized into three categories: (i) the citation count used in h-index computation, (ii) excess citation count and (iii) the citation count not used in the h-index computation (tail-citations).  \cite{bornmann2010h} referred to all those categories as the center (the h-index citation count), upper (the excess citation count) and the lower (the tail articles citation count). Based on this distribution of citation count in citation curve, the author proposed three different measures to evaluate the scientific impact of scholars:  \emph{$h^{2}$-lower, $h^{2}$-center and $h^{2}$-upper}.

These are defined as:
\begin{eqnarray}
h^{2} upper=\frac{\sum_{k=1}^{h}(cit_{k}-h)}{\sum_{k=1}^{m}cit_{k}} \times 100  = \frac{e^{2}}{\sum_{k=1}^{m}cit_{k}} \times 100
\end{eqnarray}
\begin{equation}
h^{2} center=\frac{h \times h}{\sum_{k=1}^{m}cit_{k}} \times 100
\end{equation}
\begin{equation}
h^{2} lower=\frac{\sum_{k=h+1}^{m}(cit_{k}-h)}{\sum_{k=1}^{m}cit_{k}} \times 100
\end{equation}
where,\\
$h$ is the h-index,\\ 
$cit_{k}$ is the citation count of the $k^{th}$ article,\\
$e^{2}$ is the excess citation, and\\
$m$ is the total number of articles. 

The main objective of this index is to assess the scientific impact of scholars at three different levels, which resolve the excess and tail-citation issue of h-index. A scholar having  high value in the upper section shows the high impact of the scholar and is called as perfectionist scientist. While a scholar having a low value in the upper part and high value in the lower part shows the low impact of the scholar (huge number of publication with less citation). A scholar having high center value shows the  average impact of the scholar in the community. Instead of three different measures, the only one measure should be used in the scientific assessment of scholars \cite{zhang2013h}. To do in this way, \cite{zhang2013h} proposed an improved version of $h^{2}$-lower,$h^{2}$-center and $h^{2}$-upper indices called {$h^{'}$-index}. The $h^{'}$-index of a scholar is defined as:
\begin{equation}
h^{'} = Rh=\frac{eh}{t}
\end{equation}
where $R$ represents the head-tail ratio of e and t-index. 

\cite{fred2010probing} proposed the \emph{k-index} that covers the core-tail citation ratio of a scholar. It is also called the tail-core ratio. Formally,
\begin{equation}
k=\frac{(Cit_{all}/Pub_{count})}{(Cit_{T}/Cit_{H})}
\end{equation} 
where,\\
$Cit_{all}$ is the total citation count,\\
$Pub_{count}$ is the total publication count,\\
$Cit_{T}$ is the total  citation count of h-tail articles, and\\
$Cit_{H}$ is the total citation count of h-core articles. 

The k-index of a scholar at time $t$ is defined as:
\begin{equation}
k(t)=\frac{(Cit_{all}(t)/Pub_{count}(t))}{(Cit_{T}(t)/Cit_{H}(t)}
\end{equation}
After a slight rearrangement, the above can be defined as:
\begin{equation}
k(t)=\frac{(Cit_{all}(t)Cit_{h}(t))}{Pub_{all}(t)(Cit_{T}(t)-Cit_{H}(t))}
\end{equation}
This index suffers with divide by zero problem when h-tail is zero and also it gives an average impact of scholars\cite{rousseau2011simple}. To overcome the divide by zero problem of k-index, \cite{chen2013probe} modified the k-index as  the $k^{'}$-index:
\begin{equation}
k^{'}=\frac{Cit_{all}-pub_{count}}{Cit_{T}-Cit_{H}}
\end{equation}

The \emph{Two sided h-index} by \cite{garcia2012extension} considers the core-tail ratio and states that the citation curves of two different scholars having equal or similar h-index value are different  because of the variation of citation count in their h-core and h-tail publications.The two sided h-index of a scholar for length $m$ is defined as:
\begin{equation}
h \pm m=(h_{-m}, ........,h_{-1}, h_{0}, h_{1}, .........., h_{m}) 
\end{equation}
where, \\$h_{0}$ represents the original h-index, \\the negative script represents the citations of the h-core articles with consecutive square up the top of the citation curve, and \\the positive script represents the citations of the h-tail articles with consecutive square out of the tail of citation curve. 

The negative scripted element  is calculated by finding the highest rank $R$ in h-core for $R\leq h_{m-1}$ satisfying $cit_{R} \geq R + \sum_{k=0}^{m-1} h_{-k}$.   Then $h_{-m} = R$  and the positive subscripted element is calculated by finding the largest rank $R$ for $\sum_{k=0}^{m-1}h_{k}+1\leq R \leq N$  satisfying  $cit_{R} \geq R-\sum_{k=0}^{m-1} h_{k}$, then $h_{m}=R-\sum_{k=0}^{m-1}h_{k}$.

In this subsection, we have discussed all those indices that estimate the scientific impact of scholars based on the ratio of h-core and h-tail articles citation count. But most of the indices are very difficult to compute when compared with the h-index.

\subsubsection{Based on improvement of h-index to higher value}\label{sec:Based on improvement of h-index to higher value}
The h-index considers only a few highly influential articles and can be limited due to the complete ignorance of the citations of the h-tail articles. Ignoring h-tail articles is not good because some of the articles have equal or little less citation counts than the h-index value. Such h-tail articles are obviously important in the improvement of next higher h-index. Motivated by this, \cite{ruane2008rational} proposed a new variant of h-index named \emph{rational h-index ($h_{rat}$)}. It is continuous in nature and not only considers the number of significant articles (h-index), but also the fractional progress of the scientific impact to reach the next higher value of h-index. Formally,  it is defined as:
\begin{equation}
h_{rat}=h+1-\frac{k}{2h+1}
\end{equation} 
where, $h$ is the h-index and $k$ is the number of citations required to reach $h+1$ h-index value. 

The maximum number of citations required to reach the next higher h-index is 2h+1. This index is almost similar to the h-index and the only difference is the fractional part of the index value. 

Another similar approach has been discussed by \cite{guns2009real} called \emph{real h-index}. Mathematically, the real h-index ($h_{r}$) is defined as:
\begin{equation}
h_{r}=\frac{(h+1)cit_{h}-h.cit_{h+1}}{1-cit_{h+1}+cit_{h}}
\end{equation}
where, h is the h-index and $cit_{h}$ is the citation count of $h^{th}$ article.

\cite{wu2010w} proposed the w(q)-index, where $q$ is the minimum number of additional citations required to enhance the w-index to w+1. Theoretically, it is almost similar to rational h-index, but the improvement is sought of w-index and not to the h-index. Further, it is not restricted to only w+1,  w+2, w+3 and many more also possible. Apart from these indices, \cite{nair2012stochastic} used the probability theory on the total minimum citation count required to reach the next level. They proposed the \emph{stochastic h-index}.  The stochastic h-index ($h_{s}$) of a scholar is defined as:
\begin{equation}
h_{s}=h+P
\end{equation}
where, $P$ is the probability of articles to become a part of h-core based on current publication to increase the index value to the next within a specific time period (e.g., within a year). Mathematically, it is defined as:
\begin{equation}
h_{s}=h_{0}+p=h_{0}+1-r=h_{0}+1-\sum_{p=0}^{m}r_{p}
\end{equation}
where, $r_{p}$ is the probability of the $p^{th}$ article from h-tail with citation count at least $h_{0}$ +1 within one time unit  as computed based on Poisson distribution with distribution rate $\lambda_{p}$. It is defined as:
\begin{equation}
\lambda_{p}=\frac{cit_{p}}{Y_{c}-Y_{pub}(p) +1}
\end{equation}
where, $Y_{c}$ and $Y_{pub} $ represent the current and the publication year of $p^{th}$ article respectively. Then, $r_{p}$ is defined as:
\begin{equation}
r_{p}=1-P_{k} =P(cit_{k} \leq ct_{k})
\end{equation}
where, $P(cit_{k} \leq ct_{k})$ represents the probability of additional citation count to reach the next h-index. 

In this subsection, we have discussed all those indices that try to gauge how the index value will increase to the next higher value in near future. Each of the indices gives its own mechanism to improve the index value to the next higher value. 

\subsubsection{Based on variants of citation process}\label{sec:Based on variants of citation process}
The impact of an article is measured in terms of the citation count, which refers to all those articles that cite the given article. However, it seems that a single article is cited by multiple articles of a scholar. Hence, instead of the total citation count, the total number of unique citers (readers) \cite{ajiferuke2010citer} can be an effective alternative to assess the scientific impact of an article. Motivated by this, \cite{isola2009citer} used the total number of unique citers to assess the scientific impact of an article and proposed a new index called \emph{ch-index}.  Formally, it is defined as: 

\emph{``The ch-index of a scholar is k if k of his/her articles have at least k citers each."}

To compute the ch-index, first we calculate the citers of every article.  Then, we sort the publications in the descending order of their citer-count. The ch-index is the largest rank corresponding to their citer count. But one of the main difficulty with the ch-index is to find the total number of citers of an article because none of the publications databases have citer information. So, it is a very difficult task to do.

\cite{katsaros2009f} give an alternative way to measure the total impact of an article by the number of distinct cited scholars called the \emph{coterminal citation count} and proposed a new index called \emph{f-index}. The f-index of a scholar is defined as :

\emph{``The f-index of a scholar is k if k of his/her articles have at least k coterminal citation each."}

The coterminal citation count of an article is the dot product of m-dimensional quantities and $S$. The quantities of an article is the ratio of the cardinalities of the set of authors of the citing article and the total number of distinct authors of all the cited article. $S$ is the vector containing ${m, m-1, m-2, ....., 1}$. Where $m$ is the total number of citing articles. The series of the set of  the author is defined as:
\begin{equation}
S_{K}^{A}=\{A_{p}: author~ A_{p}~ present~ precisely~ in~ K^{th} ~citing~ article \}
\end{equation}  
The quantities of citing article are defined as:
\begin{equation}
F_{k}^{A}=\frac{|S_{K}^{A}|}{Total~distinct~citing~author} 
\end{equation}
For calculation of the f-index, we first calculate the co-terminal citations of all article, then arrange all articles in descending order of their coterminal citation count. Then the f-index of a scholar is the largest rank in which the coterminal citation count is greater than or equal to the rank. This index tries to remove the author reparation from the citation process. But this index is a little bit difficult to compute.    

An article is written with the help of other articles it refers to. In this way, the impact of an article can be based on the impact of the refereed articles. If an article refers to the highly influential article, then it has good chances to gain more higher impact in future. Along these lines, \cite{schubert2008using} proposed the \emph{Single publication h-index}. The single publication h-index is based on highly cited referred articles. For calculation of single publication h-index, first we arrange the referred articles in the descending order of their citation counts. The single publication h-index is the h-index value from the reference list. Using the single publication h-index, \cite{egghe2011single} proposed the \emph{Indirect h-index}.  For the calculation of the indirect h-index, we first calculate the single publication h-index of all publications. Then, we arrange all the publications in the descending order of their single publication h-index. The indirect h-index of a scholar is defined as: 

\emph{``The indirect h-index of a scholar is m, if m of his/her articles have at least m single publication h-index each."}

\cite{egghe2011single1} discussed the indirect h-index and improved this index based on Lotka's principle:
\begin{equation}
IH=h^{\frac{\alpha}{1+\alpha^{'}(\alpha -1)}}
\end{equation}
where $\alpha$ and $\alpha^{'}$ are Lotka parameters defined as $\alpha$ $>$ 1 and $\alpha^{'}$ is:
\begin{equation}
\alpha^{'}=\frac{ln(h)}{ln(H_{h})}
\end{equation} 
Another index based on single publication h-index is the \emph{average h-index}: 

\emph{``The average h-index of a scholar is the average of the single publication h-index of h-core articles."}\\
Mathematically, it is defined as:
\begin{equation}
\overline{H}=\frac{1}{h}\sum_{k=1}^{h} H_{k}
\end{equation}

But the single publication h-index is very difficult to calculate because none of the publications databases has any information related to reference citation count. 

The citation count of an article is influenced by the self-citation count, citations from co-authors articles and from institutional colleagues. All those types of citations are treated as self-citation. Instead of the total number of citation count, the total number of international recognition is a much better way to know the scientific impact of an article. Based on this international recognition \cite{kosmulski2010hirsch1} proposed the \emph{$h_{int}$}-index. The $h_{int}$-index of a scholar is defined as:

\emph{``The $h_{int}$-index of a scholar is k, if k of his/her articles have at least k number of international recognition each."}

The international recognition of an article is the sum of all those articles that refer to other country scholars' publications. To compute the $h_{int}$-index, first calculate the total number of international recognition of each article. Then sort the publication data into the descending order of their total number of international recognition. The $h_{int}$-index of a scholar is the largest rank in which the corresponding international recognition value is greater or equal than the rank. 

In a research article, it seems that the some of the references are cited at once while some are cited multiple times. In both cases, the citation count of an article remains the same. So, instead of total citation count, the total number of the distinct author or the total number of international recognition, the total number of citation mention could be used an effective alternative to assess the impact of an article. To incorporate the citation mentioned, \cite{wan2014wl} proposed a new variant of h-index called \emph{wl-index}.

\emph{``The wl-index of a scholar is the largest integer k, if k of his/her articles have at least k citation mentioned each."}

In this subsection, we have discussed all those indices that are based on alternative citation processes of an article. Every index gives its own benchmark to know the scientific impact of articles and scholars.

\subsubsection{Miscellaneous indices}\label{sec:Miscellaneous indices}
Many other indices that have not been covered under the above mentioned categories have been covered in this section. They are also important in scientific assessment of scholars. All such indices mentioned in this section are trivial to calculate than the other indices mentioned previously.

\cite{van2008generalizing} mentioned that the h-index definition is somewhat arbitrary \cite{lehmann2006measures,lehmann2008quantitative} being based on the number of highly cited articles.  A single arbitrary number is not fair in assessing the scientific impact of scholars and while comparing two different scholars. To overcome this limitation \cite{van2008generalizing} proposed a new index called \emph{$h_{\alpha}$-index}, which depends on a parameter $\alpha$.

\emph{``The $h_{\alpha}$-index of a scholar is k, if k of his/her articles have at least $\alpha k$ citations each."}

where $\alpha >1$.  

For example, let scholar $X$ publish 20 articles with 20 citations each. Let scholar $Y$ publish 10 articles with 100 citations each. Then, the h-index of scholar $X$ is 20, while for scholar $Y$, it is 10. But the scientific impact of author $Y$ is more than author $X$. If the value of $\alpha$ is 10, then the $h_{\alpha}$  is 2  and 10 respectively  for authors X and Y. If we set $\alpha$ = 5, then the $h_{\alpha}$ is 4 and 10 respectively for authors $X$ and $Y$. Finally, we can say that the $h_{\alpha}$-index gives more fine grained value than the other h-types indices. But, the main limitations of this index is the determination of the value of $\alpha$.

The citation based indices are designed based on the number of articles having significant amount of influence. The articles being considered  are either journal articles, conference articles or book chapters. Further, all of the articles are treated equally for the scientific assessment. But every type of publications has its own impact and definitions. So, we can not treat all equally. To overcome this limitation, \cite{vinkler2009pi} proposed a new index based on an elite set of articles called \emph{$\pi$-index}. The elite set of highly cited journal article is the square root of total article count  ($A_{\pi}=\sqrt{A}$). The $\pi$-index of a scholar is the $100^{th}$ of the number of citation count of an elite set of highly cited journal articles that are sorted in descending order of their citation count. Mathematically, it is defined as:
\begin{equation}
\pi-index =0.01\times cit(A_{\pi})
\end{equation}
where, $cit(A_{\pi})$ is the citation count of an elite set of articles.

\cite{bornmann2010citation,schubert1986mean,van2005signals} use the speed of the citation count of an article to evaluate the scientific impact of scholars. The speed of the citation count is measured at which the first citation count of an article is gained. For the calculation of the citation speed index of a scholar, first the total number of months after the article earns its first citation count is calculated for each of the publications. Then,  the publication list is arranged in the descending order of their month count. The citation speed index of a scholar is then defined as:

\emph{``The citation speed index of a scholar is k, if k of his/her articles got first citation at least k months ago."}\\
Mathematically, it is defined as:
\begin{equation}
SI=\underset{k}{max}(Month_{k} \geq k)
\end{equation}
where $k$ is the largest rank and $Month_{k}$ is the total number of months in which the first citation count of $k^{th}$ article was earned.

This index penalizes all those articles that have earned citations earlier.

Instead of citation count of highly cited articles, the citation count of all those articles that have a significant amount of influence in the proportion of collaboration distance between the cited and the citing scholars, have been used for the scientific assessment of scholars.  Following this, \cite{domingobibliometric,bras2011bibliometric} considered the impact of collaboration distance between the cited and the citing scholars and proposed the \emph{c-index}. The collaboration distance between cited and citing scholar is measured in two categories: (i) Classical collaboration distance and (ii) Refined collaboration distance. 

The classical collaboration distance is the total number of scholars who are present between the citing and the cited scholars. This collaboration distance requires at least one collaboration between any member of the citing and the cited scholars. For example, let scholar A1, A2, and A3 be the citing scholars, and scholar A7, A5, and A4  be the cited scholars. Let one collaboration between A2 and A5 be present, hence, the collaboration distance between scholar A1 \& A4 is 4. The collaboration distance between cited and citing article is 1 because only one collaboration distance is present between these two articles.

In the refined collaboration distance, the distance between two scholars is the total collaboration path length, defined as:
\begin{equation}
\sum_{r=0}^{m-1}\frac{1}{|pub(A_{r}) \cap pub(A_{r+1})|}
\end{equation}
where, $m$ is the total number of scholars between the cited and the citing scholars. 

Based on these two types of collaboration distances, the c-index \cite{domingobibliometric,bras2011bibliometric} is defined as:
\begin{equation}
C_{\alpha}=max(min(\alpha k, Qt(p_{k}))) : k \in \{1,2,...,m\}
\end{equation} 
where, $\alpha$ is the slope in citation graph and $Qt(p_{k})$ is quality function of $p_{k}$.

In the case of the c-index of an article, the quality function $Qt(p_{k})$ is the collaboration distance between the cited and the citing article. In case of the c-index of a scholar, the quality function $Qt(p_{k})$ is the citation distance between the scholars of the citing and the cited articles. If a scholar published all papers by self without any co-author, then the collaboration distance is $\infty$ and the c-index is $\alpha$m.

One another variant of the h-index is based on the number of success articles \cite{moed2010measuring} and was defined by \cite{kosmulski2011successful}. If the total citation count of an article is more than the total number of references, then the article is called a success article. The score (SC) of the success of an article is defined as:
\begin{equation}
SC_{A}=\begin{cases}
1, &\textrm{if $cit_{A} > RC_{A}$} \\
0, ~~~ &\textrm{Otherwise}
\end{cases}
\end{equation}
where,\\
$SC_{A}$ represents the score of article $A$,\\
$cit_{A}$ is the citation count of article $A$, and\\
the $RC_{A}$ is the total reference count of article $A$. 

The success article penalizes all review articles because the review articles contain a huge number of references. This method also penalizes the newly published articles. Suppose an article earns 5 citation counts and contains 4 references, then this article called a success article.  But an article that earns 100 citation count and contains 120 references, does not count as a success article. To overcome this limitation, a new index was proposed in reference \cite{kosmulski2011successful} to calculate the score of an article. Formally, it is defined as:
\begin{equation}
SC_{A}=\begin{cases}
\frac{cit_{A}}{RC_{A}+1}, &\textrm{if $cit_{A} > RC_{A}$} \\
0, ~~~ &\textrm{Otherwise}
\end{cases}
\end{equation}
The total success score of a scholar is the sum of the success score of all articles.
\begin{equation}
NSA=\sum_{A=1}^{m}SC_{A}
\end{equation}
where, \\
$NSA$ represents the number of success articles,\\
$m$ is the total number of articles published by the scholar, and\\
$SC_{A}$ is the success score of an article.  

In case of multi-authored articles, the success score of an article is defined as:
\begin{equation}
SC_{A}=\begin{cases}
\frac{1}{M_{A}}, &\textrm{if $cit_{A} > RC_{A}$} \\
0, ~~~ &\textrm{Otherwise}
\end{cases}
\end{equation}
where, $M_{A}$ is the total number of authors in article $A$. 

The citation of an article increases over time, so age of an article is an important factor to evaluate the  scientific impact of an article. Based on this factor the score of an article is defined as:
\begin{equation}
SC_{A}=\begin{cases}
\frac{1}{CY + 1 - PY }, &\textrm{if $cit_{A} > RC_{A}$} \\
0, ~~~ &\textrm{Otherwise}
\end{cases}
\end{equation}
where, $CY$ and $PY$ represent current year and publication year respectively. 

Further, \cite{franceschini2012success} discussed the success paper based index and mentioned that it also suffers with excess citation count (as discussed in \cite{franceschini2010citation}). To overcome this limitation of NSP (number of success paper) based indicator, \cite{franceschini2012success} proposed a new index called \emph{Success-index}. The success-index of a scholar is the sum of the score of all articles. Formally, it is defined as:
\begin{equation}
Success-index(A)=\sum_{k=1}^{m}SC_{k}
\end{equation}
where, $SC_{k}$ is  score of the $k^{th}$ success article (\cite{kosmulski2011successful}).  

An article consists of two components: (i) article citation count and  (ii) publication impact factor. However, generally only the number of citations is used and the impact factor of publications is completely ignored.  Further, the citation count of an article is categorized into the following categories:
\begin{enumerate}
	\item A citation by scholars who have never worked with the cited scholar.
	\item A citation by scholars who have never worked with the cited scholar, but have collaborated with an intermediate scholar(s).
	\item A citation by the scholars who are co-authors in another publication.
	\item A citation by the co-author except for researchers whom the individual scientific activity is established. 
	\item Self-citation.  
\end{enumerate}
By using the impact of different types of citation count and the publication impact factor, \cite{mikhailov2012new} proposed a new index named \emph{Summary Citation Index}.  The value of the reference is determined by the impact factor of publication venue  \cite{erren2016research}. The value coefficient ($\phi$) is introduced to assign weight to the impact of different references. In all those five citation sources discussed above, the first one is more valuable than others. The value coefficient is 1 for the first type of citation and 0.9, 0.75, 0.5 \& 0.25 respectively for other types of citations. The product of impact factor and the value coefficient is distributed equally to all scholars in multi-authored articles. The citation count of an article for a scholar is defined as:
\begin{equation}
OC=\sum_{p=1}^{m}\frac{\Phi(\phi_{p}IF_{p})}{A_{p}}
\end{equation}
where, \\
$OC$ represents the summary citation count,\\
$\Phi$ represents correction factor,\\
$\phi_{p}$ \& $IF_{p}$ are  the value coefficient and impact factor of $p^{th}$ article, \\
$A_{p}$ represents the number of scholars in $p^{th}$ article, and\\
$m$ is the total number of articles.

The Edition Citation (EC) with additional coefficient $a$ is defined as:
\begin{equation}
EC=a\sum_{p=1}^{m}\frac{\Phi(IF_{p})}{A_{p}}
\end{equation}
The value of $a$ is typically set as 100. 

Finally, the author  mentioned that $OC$ and $EC$ are additive values.  By using the citation count of an article and the impact factor of the publication venue, the  summary citation (SC)-index is defined as  the sum of $OC$ and $EC$ parameters. Formally, 
\begin{equation}
SC= OC + EC = \sum_{p=1}^{m}\frac{\Phi(\phi_{p}IF_{p})}{A_{p}}+a\sum_{p=1}^{m}\frac{\Phi(IF_{p})}{A_{p}}
\end{equation}

Generally, all of the indices produce only a single number to evaluate the scientific impact of scholars. In the process, they  fail (i) to cover the yearly impact of scholars, and (ii) to compare the scientific impact of scholars. Motivated by this, \cite{liu2014empirical} discussed the h-index sequences and proposed the L-sequence. The  L-sequence considers the entire research career of a scholar to determine the scientific impact. To define L-sequence, consider a scholar who has published $k$ articles in his/her career. Let the first publication year be $y_{1}$ and the current year be $y_{c}$. Then, the L-sequence of an author for $n^{th}$ year, denoted $L_{n}$, is the h-index value computed on the basis of the citation counts of all the publications received in the $n^{th}$ year. Thus, the L-sequence of the author is  $L_{y_{1}}, L_{y_{1}+1}, ...............L_{y_{c}}$.  

Apart from above indices, \cite{bertoli2015formula} discussed the mathematical model of h-index. Mathematical model of h-index for a scholar is defined as:
\begin{eqnarray}
\tilde{h_{w}}= \tilde{h_{w}}(M, Cit, Cit_{max}) = \left[\frac{W((\frac{M-1}{(1-\tilde{k}^{-1})}).log\frac{1}{(1-\tilde{k}^{-1})})}{log(1/(1-\tilde{k}^{-1}))}\right]
\end{eqnarray}
where, $w$ indicates Lambert function and $\tilde{k}$ represents trimmed mean, which is defined as:
\begin{equation}
\tilde{k}=\frac{\tilde{Cit}}{(M-1)}
\end{equation} 
Here, $M$ represents the total publication count and $\tilde{Cit}$ = Cit-$cit_{max}$.

In this subsection, we have discussed all those indices which are a little bit difficult to compute because of them being based on complex mathematical models. Finally, we have concluded that in the field of scientometrics and bibliometrics, several indices covers several things to asses the scientific impact of scholars, but there is still an opportunity to improve the evaluation process of the scientific impact of scholars. Several indices complemented the h-index in the context of the total number of core items, whereas several others cover the impact of co-authors in scientific assessment. Apart from the core items and the number of co-authors, several indices consider the excess citation count of core articles, some consider the total number of cited articles, the total career of scholars, and more, to enhance the scientific assessment of scholars. A huge number of indices use mathematical expressions to share credit among co-authors, but simply the use of mathematical formulas to express the contribution of scholars is not justifiable. So, in this regard, there is scope for further research. Lastly, only a limited number of articles have addressed the impact of excess citation count.  

Form the above study, we found that a limited number of research have been conducted to account for the importance of excess citation count, tail-articles citation count and the publications consistency. These all have a significance in the scientific assessment of scholars. Our research is solely based on the above mentioned issues.    The summary of the index based on other variants shown in the table \ref{tab:Summary of Index based other variants}.
\begin{table}[!h]
	\renewcommand{\arraystretch}{1.3}
	\caption{Summary of Index based other variants}
	\label{tab:Summary of Index based other variants}
	\centering
	\begin{tabular}{lp{9.2cm}p{2.0cm}}
		\hline
		\textbf{Index}     &\textbf{Definition}    &\textbf{Publication}\\
		\hline
		&\textbf{Based on core-tail ratio} &\\\hline
		$h^{2}$ upper-index & $h^{2} upper=\frac{\sum_{k=1}^{h}(cit_{k}-h)}{\sum_{k=1}^{m}cit_{k}} \times 100 $ &\cite{bornmann2010h}\\
		
		$h^{2}$ center-index &$h^{2} center=\frac{h \times h}{\sum_{k=1}^{m}cit_{k}} \times 100 $ &\cite{bornmann2010h}\\
		
		$h^{2}$ lower-index &$h^{2} lower=\frac{\sum_{k=h+1}^{m}(cit_{k}-h)}{\sum_{k=1}^{m}cit_{k}} \times 100$ &\cite{bornmann2010h}\\
		
		$h^{'}$-index &$h^{'} = Rh=\frac{eh}{t}$ Where R represents head-tail ratio by e and t-index. &\cite{zhang2013h}\\
		
		k-index & $k=\frac{(cit_{all}/pub_{count})}{(cit_{T}/Cit_{H})}$
		$cit_{all}$ is total citation count, $pub_{count}$ is total publication count, $cit_{T}$ total citation count of h-tail papers and $cit_{H}$ is the total citation count of h-core papers. &\cite{fred2010probing}\\
		
		$k^{'}$-index & $k^{'}=\frac{cit-pub}{Cit_{T}-cit_{H}}$
		&\cite{chen2013probe}\\
		
		Two sided h-index & \emph{The two sided h-index of a scholar for length $m$ is defined as:
		$h \pm m=(h_{-m}, ........,h_{-1}, h_{0}, h_{1}, .........., h_{m}) $
		where $h_{0}$ represents the original h-index, the negative script represents the citations of the h-core articles with consecutive square up the top of the citation curve, and the positive script represents the citations of the h-tail articles with consecutive square out of the tail of citation curve. } &\cite{garcia2012extension}\\\hline
		
		& \textbf{Based on improvement of h-index to higher value}&\\\hline		
		
		Rational h-index ($h_{rat}$) &\emph{The rational h-index of an author is the h and the number of additional citation required to reach next h-index(h+1)}, $h_{rat}=h+1-\frac{k}{2h+1}$, Where h is the h-index and k is the number of citation count required to reach h+1. The maximum number of citation required to reach next h-index is 2h+1. &\cite{ruane2008rational}\\
		
	 	Real h-index & $h_{r}=\frac{(h+1)cit_{h}-h.cit_{h+1}}{1-cit_{h+1}+cit_{h}}$ where, h is the h-index and $cit_{h}$ is the citation count of $h^{th}$ article. &\cite{guns2009real}\\
		
		\hline
		\end{tabular}
	\end{table}
\begin{table}[]
\addtocounter{table}{-1}
\renewcommand{\arraystretch}{1.2}
\footnotesize
\caption{Summary of Index based other variants Continue..}
\centering
\begin{tabular}{lp{9.2cm}p{2.0cm}}
\hline
\textbf{Index}     &\textbf{Definition}    &\textbf{Publication}\\
\hline
		 Stochastic h-index ($h_{s}$) & $h_{s}=h+P$, where, $P$ is the probability of articles to become a part of h-core based on current publication to increase the index value to the next within a specific time period (e.g., within a year). &\cite{wu2010w}\\\hline
		 
		 &\textbf{Based on variants of citation process}&\\\hline

		ch-index &\emph{``The ch-index of a scholar is k if k of his/her articles have at least k citers each."} &\cite{isola2009citer}\\

		f-index &\emph{``The f-index of a scholar is k if k of his/her articles have at least k coterminal citation each."} &\cite{katsaros2009f}\\
	
		Indirect h-index &\emph{``The indirect h-index of a scholar is m, if m of his/her articles have at least m single publication h-index each."} &\cite{egghe2011single}\\
		
		Average h-index &\emph{``The average h-index of a scholar is the average of the single publication h-index of h-core articles."} &\cite{egghe2011single1}\\
					
		$h_{int}$-index &\emph{``The $h_{int}$-index of a scholar is k, if k of his/her articles have at least k number of international recognition each."} &\cite{kosmulski2010hirsch1}\\
		
		wl-index &\emph{``The wl-index of a scholar is the largest integer k, if k of his/her articles have at least k citation mentioned each."} &\cite{wan2014wl}\\\hline
		
		&\textbf{Miscellaneous indices}&\\\hline

		$h_{\alpha}$-index &\emph{``The $h_{\alpha}$-index of a scholar is k, if k of his/her articles have at least $\alpha k$ citations each."} &\cite{van2008generalizing}\\
		
		$\pi-index$ &$\pi-index =0.01\times cit(A_{\pi})$, where, $cit(A_{\pi})$ is the citation count of an elite set of articles. &\cite{vinkler2009pi}\\
		
		Citation speed index &\emph{``The citation speed index of a scholar is k, if k of his/her articles got first citation at least k months ago."} &\cite{bornmann2010citation}\\
		
		c-index &$C_{\alpha}=max(min(\alpha k, Qt(p_{k}))) : k \in \{1,2,...,m\}$ &\cite{domingobibliometric}\\
		
		Success index &\emph{The success index of an author sum of the success score of all article.} ($Success-index(A)=\sum_{k=1}^{m}SC_{k}$, where $SC_{k}$ is success score of $k^{th}$ publication.) &\cite{franceschini2012success}\\
		
		Summary citation index &$SC= \sum_{p=1}^{m}\frac{\Phi(\phi_{p}IF_{p})}{A_{p}}+a\sum_{p=1}^{m}\frac{\Phi(IF_{p})}{A_{p}}$, &\cite{mikhailov2012new}\\

		$\tilde{h_{w}}$&$\tilde{h_{w}}= \tilde{h_{w}}(M, Cit, Cit_{max}) = \left[\frac{W((\frac{M-1}{(1-\tilde{k}^{-1})}).log\frac{1}{(1-\tilde{k}^{-1})})}{log(1/(1-\tilde{k}^{-1}))}\right]$, Where w indicates Lambert function and $\tilde{k}$ represents trimmed mean. &\cite{bertoli2015formula}\\
				\hline 
	\end{tabular}
\end{table}

\section{Conclusions}
In this article, we made an extensive literature review on h-index and its alternative indices. In research community h-index plays an important role to evaluate the scientific impact of individual, Institutions, college or university to grant research project as well as a promotion or award. The h-index got a lot of attention due to its simplicity.

Based on the characteristics of h-index and also to overcome the limitation of h-index,  several other indicators were proposed by eminent researchers. Some of the variants extended the properties of h-index to overcome h-index big hit problems, while some eminent researchers proposed other variants of h-index to overcome the multi-authored problem. Many others indices were proposed to overcome publication age problems, excess citation count problem, total publication count problem. Some other variants that combine the advantages of two indices to overcome shortcomings of other indices.  Several other indicators were proposed considers other parameters such as a number of citer, number of citation mentioned, L-factor, number of success papers and so on. Some of the researchers discussed the impact of self citation count and proposed some other variants of h-index for account of self citation count.

Some other variants of h-index is based on the Pagerank algorithm to assign influence of an article in the research community. 

Here we found that all indices is categories in eight categories as Complement of h-index, based on total number of author, publication age, Combination of two indices, excess citation, total publication count, other variants and account for self citation count.

In research community many other literatures discussed the h-index and its alternatives theoretically \& empirically and also discuss the advantages \& limitation of proposed variants.     

\bibliographystyle{apalike}      
\bibliography{report}                
\nocite{*}

\end{document}